\documentclass[12pt]{article}
\usepackage{amssymb,amsmath,epsfig}
\makeatletter

\@addtoreset{equation}{section}
\def\section{\@startsection {section}{1}{\z@}{-2.25ex plus -1ex minus
 -.2ex}{0.8ex plus .2ex}{\large\bf}}
\def\subsection{\@startsection{subsection}{2}{\z@}{-2.0ex plus%
 -1ex minus -.2ex}{0.4ex plus .2ex}{\bf}}

\def\Ad{\mathrm{Ad}}

\newcommand{\tr}[0]{{\rm tr}}

\newcommand{\inv}[0]{{-1}}

\newcommand{\hyp}[0]{{\mathbb{H}^2}}

\def\bv{{\mbox{\boldmath $v$}}}
\def\bw{{\mbox{\boldmath $w$}}}
\def\ba{{\mbox{\boldmath $a$}}}

\def\bx{{\mbox{\boldmath $x$}}}

\def\by{{\mbox{\boldmath $y$}}}

\def\bz{{\mbox{\boldmath $z$}}}

\def\bp{{\mbox{\boldmath $p$}}}

\def\bq{{\mbox{\boldmath $q$}}}

\def\bn{{\mbox{\boldmath $n$}}}

\newcommand{\RR}{\mathbb{R}}
\newcommand{\CC}{\mathbb{C}}
\newcommand{\MM}{\mathbb{M}}
\newcommand{\HH}{\mathbb{H}}

\def\bea{\begin{eqnarray}}
\def\eea{\end{eqnarray}}
\def\bmz{\left(\begin{array}{2,2}}
\def\emz{\end{array}\right)}
\def\bmd{\left(\begin{array}{3,3}}
\def\emd{\end{array}\right)}

\def\bpm{\begin{pmatrix}}
\def\epm{\end{pmatrix}}

\begin{document}
\parskip 6pt
\parindent 0pt

\begin{center}
\baselineskip 24 pt {\Large \bf  Cosmological measurements, time and observables in (2+1)-dimensional gravity}

\baselineskip 16 pt

\vspace{.5cm} { C.~Meusburger}\footnote{\tt  catherine.meusburger@nottingham.ac.uk}\\
School of Mathematical Sciences\\
The University of Nottingham\\
University Park\\ Nottingham NG7 2RD, United Kingdom

\vspace{0.5cm}

{25 November  2008}

\end{center}

\begin{abstract}
\noindent
We investigate the relation between  measurements  and the physical  observables for  vacuum spacetimes with compact spatial surfaces in (2+1)-gravity with vanishing cosmological constant. 
By considering an observer who emits lightrays that return to him at a later time, we obtain explicit expressions for several measurable quantities as functions on the physical phase space of the theory:
 the eigentime elapsed between the emission of a  lightray and its return to the observer,  the angles between the directions into which the light has to be emitted  to return to the observer and the relative frequencies of the lightrays at their emission and return.  This provides a framework in which conceptual questions about time, observables and measurements can be addressed. We analyse the  properties of these measurements and their geometrical interpretation and show how they allow an observer to determine the values of the Wilson loop observables that parametrise the physical phase space of (2+1)-gravity. We discuss the role of time in the theory and demonstrate that the specification of an observer with respect to the spacetime's geometry amounts to a gauge fixing procedure yielding Dirac observables.

 \end{abstract}

\section{Introduction}
\label{intro}

Gravity in (2+1) dimensions has been investigated extensively as a toy model for the quantisation of  higher-dimensional gravity, for an overview see \cite{Carlipbook, carliprev}. As the theory simplifies considerably in (2+1)-dimensions, it  becomes amenable to quantisation and thus provides a framework in which conceptual questions of quantum gravity can be investigated in a fully quantised theory. 
However, this goal is obstructed by a problem present also in (3+1) dimensions: it is difficult to relate the variables parametrising the phase space and used in quantisation to physically meaningful quantities that could be measured by an observer.  

Although (2+1)-dimensional gravity is equipped with a complete set of gauge and diffeomorphism invariant observables, the Wilson loops 
along closed curves in the spacetime, it is  currently unclear how these observables are related to realistic physical measurements performed by observers.
This hinders the application of the resulting quantum theory to concrete physical problems and complicates the interpretation even on the classical level. In particular, it is not known how to define operators with a clear physical interpretation, how the Wilson loop observables that parametrise the phase space could be reconstructed from measurements 
performed by observers and how time variables such as the observer's eigentime enter the theory.

In this paper, we address this problem for Lorentzian vacuum  spacetimes in classical (2+1)-gravity with vanishing cosmological constant. More specifically, we consider maximal globally hyperbolic vacuum spacetimes with compact genus $g\geq 2$ spatial surfaces,  which resemble the Bianchi models in (3+1) dimensions.  
We pursue an approach similar to gravitational lensing  and consider an observer  who probes the geometry of the spacetime by emitting lightrays.
Such an observer will notice that the  lightrays sent in certain directions return to him, and he can measure the amount of eigentime elapsed between the emission and reception of such a returning lightray. 
Moreover, the observer can determine the directions into which the light needs to be emitted in order to return and the angles between these directions. He also can compare the frequencies
  of the lightray at its emission and return. 

This provides us with physically meaningful measurements  that resemble the ones performed in cosmology and astrophysics. The purpose of this paper is to relate these measurements to the observables that parametrise the phase space of the theory and serve as the fundamental building blocks in its quantisation.  More specifically, we resolve the following issues:
\begin{enumerate}
\item We derive explicit expression for these measurements in terms of the fundamental observables of (2+1)-dimensional gravity, the holonomies along closed curves in the spacetime and the associated Wilson loop observables.

\item We discuss  their physical properties, analyse their geometrical interpretation and show how they encode the geometry of the underlying spacetime.

\item We demonstrate how an observer can reconstruct the values of the holonomies and Wilson loop observables and hence the physical state of the spacetime from these measurements. 

\item We give a careful discussion of the conceptual issues of quantum gravity that manifest themselves in this description. In particular, we discuss the role of partial and complete observables and show that   specifying an observer with respect to the geometry of the spacetime amounts  to a gauge fixing procedure. 

\item We investigate the role of time in the theory. In particular, we find that the observer's eigentime plays the role of an additional parameter that relates his measurements to the gauge and diffeomorphism invariant observables parametrising the phase space. 
\end{enumerate}

Together, these results define a set of physical quantities   that could be measured by an observer. These quantities determine the spacetime's geometry uniquely and are given  explicitly as functions on the physical phase space of the theory.  This provides a framework in which conceptual questions about time, observables and the phase space can be addressed. 
In particular, it offers the prospect of investigating the associated operators in the quantum theory and of clarifying fundamental conceptual questions of quantum gravity.

The paper is structured as follows. In Sect.~\ref{spacetimesect} we give an overview of the geometrical properties of flat maximal globally hyperbolic vacuum spacetimes of topology $\RR^+\times S_g$, where $S_g$ is an oriented two-surface of genus $g\geq2$. Following the presentation in \cite{mess, bb}, we review the description of such spacetimes as quotients of regions in Minkowski space by the action of cocompact Fuchsian groups and the construction of evolving spacetimes via grafting.

Sect.~\ref{obsmotiv} contains a brief discussion of the conceptual questions in classical and quantum
gravity that are associated with observables, time and physical measurements. We motivate and summarise the central idea of this paper - to consider observers that measure the geometry of the spacetime and determine the values of the physical observables via returning lightrays - and discuss its relation to gravitational lensing.

In Sect.~\ref{messect} we derive the main results of our paper. We consider three realistic physical quantities that could be measured by an observer in the spacetime: the eigentime elapsed between the emission and reception of a returning lightray, the directions in which an observer needs to send light in order to have it return to him and the angles between these directions as well as the relative shift in frequency between the emitted and the returning lightray. We derive explicit expressions for these quantities as functions of the observer's eigentime, his worldline and of the observables that parametrise the physical phase space of the theory. We discuss their physical interpretation and show how they encode the  spacetime's geometry.

In Sect.~\ref{physdiscsect} we discuss our results with respect to the conceptual questions of classical and quantum gravity outlined in Sect.~\ref{obsmotiv}. We show that they provide 
a framework in which these questions can be addressed explicitly and concretely. 
 In particular, we 
demonstrate that the measurements in Sect.~\ref{messect} are related to the gauge invariant observables of the theory via the  specification of an observer with respect to the geometry of the spacetime, which can be viewed as a gauge fixing procedure.  We discuss the role of time in the theory and give an explicit prescription through which the observer can determine the physical observables from his measurements associated with  returning lightrays. 

Sect.~\ref{outlook} contains our outlook and conclusions. The appendix summarises  facts and definitions from two-dimensional hyperbolic geometry and the theory of cocompact Fuchsian groups.


\section{Vacuum spacetimes in (2+1)-gravity}
\label{spacetimesect}

\subsection{Definitions and notation}

\label{defnotsect}

Throughout the paper, we use Einstein's summation convention. Indices run from 0 to 2 and are raised and lowered with the (2+1)-dimensional Minkowski metric $\eta=\text{diag}(-1,1,1)$. We use the notations
$\bx^2=\eta(\bx,\bx)=x_ax^a$ and  $\bx\cdot\by=\eta(\bx,\by)=x_ay^a$. 
For $\bn\in\RR^3$ timelike ($\bn^2<0$) or spacelike ($\bn^2>0$), we denote by  $\hat\bn$  the associated unit vector satisfying, respectively, $\hat \bn^2=-1$ and $\hat\bn^2=1$.
We write $\bn^\bot=\{\by\in\RR^3\,|\,\bn\cdot\by=0\}$ for the orthogonal complement of $\bn\in\RR^3$
 and  $\Pi_{\bn^\bot}$ for  the projection on $\bn^\bot$
 \begin{align}
\label{tangproj}
\Pi_{\bn^\bot}(\bv)=\begin{cases} 
\bv-(\bv\hat\bn)\bn & \text{for}\; \bn^2>0 \;(\bn \;\text{spacelike})\\
\bv+(\bv\hat\bn)\hat\bn & \text{for}\;\bn^2<0 \;( \bn\;\text{timelike})
\end{cases}\qquad\forall \bv\in\RR^3.
\end{align}
The proper orthochronous Lorentz group in three dimensions is the group $SO(2,1)_0^+\cong PSL(2,\RR)\cong PSU(1,1)$. We fix a set of generators $J_a$, $a=0,1,2$, of its Lie algebra in terms of which the Lie bracket takes the form
\begin{align}
\label{liebracket}
[J_a,J_b]=\epsilon_{abc}J^c,
\end{align}
where $\epsilon$ is the totally anti-symmetric tensor in three-dimensions with the convention  $\epsilon_{012}=-\epsilon^{012}=1$.  A set of $\mathfrak{su}(1,1)$-matrices satisfying these relations is given by
\begin{align}
\label{jmatrix} &J_0=\tfrac{1}{2}\left(\begin{array}{cc} i &
0\\0 &
-i\end{array}\right) & &J_1=\tfrac{1}{2}\left(\begin{array}{cc} 0 & -i\\
i
& 0\end{array}\right) & &J_2=\tfrac{1}{2}\left(\begin{array}{cc} 0 & 1\\
1 & 0\end{array}\right).
\end{align}
Using this representation, we obtain for the exponential map $\exp: \mathfrak{su}(1,1)\rightarrow SU(1,1)$
\begin{align}
\label{su11rep}
\exp(n^b J_b)=\begin{cases}\cosh\tfrac {|\bn|} 2\; 1+2 \sinh \tfrac {|\bn|} 2 \hat n^bJ_b & \text{for}\; \bn^2>0\;(\bn\,\text{spacelike})\\
\cos\tfrac {|\bn|} 2\; 1+2 \sin \tfrac {|\bn|} 2 \hat n^bJ_b & \text{for}\; \bn^2<0\; (\bn\,\text{timelike})\\
1+n^bJ_b& \text{for}\; \bn^2=0\; (\bn\,\text{lightlike}).
\end{cases}
\end{align}
Note that this map  is neither injective nor surjective. However, the induced map $\exp: \mathfrak{su}(1,1)\rightarrow PSU(1,1)=SU(1,1)/\mathbb Z_2$ is surjective and elements of the proper orthochronous Lorentz group can therefore be parametrised via \eqref{su11rep}. Elements of $PSU(1,1)\cong SO^+_0(2,1)$ are called hyperbolic, parabolic and elliptic, respectively, if the vector $\bn$ in \eqref{su11rep} is spacelike ($\bn^2>0$), lightlike ($\bn^2=0$) and timelike ($\bn^2<0$).

The standard representation of the (2+1)-dimensional proper orthochronous Lorentz group on $\RR^3$ by $SO(2,1)_0^+$-matrices  agrees with the adjoint action of $PSU(1,1)$ on its Lie algebra $\mathfrak{su}(1,1)\cong \mathfrak{sl}(2,\RR)\cong\mathfrak{so}(2,1)$ and will be denoted by $\Ad(v)$ in the following
\begin{align}
\label{adjact}
v\cdot J_a\cdot v^\inv=\Ad(v)^b_{\;\;a}J_b\qquad\forall v\in SU(1,1).
\end{align}
Using \eqref{su11rep} we find that the action of this representation on vectors $\bx\in\RR^3$ is given by
\begin{align}
\label{adrep}
&\Ad(\exp(n^bJ_b))\bx
&=\begin{cases}\cosh |\bn|(\bx-(\hat\bn\bx)\hat\bn)+(\hat\bn\bx)\hat\bn+\sinh |\bn| \hat\bn\wedge\bx & \text{for}\;\bn^2>0\\
\cos |\bn|(\bx+(\hat\bn\bx)\hat\bn)+(\hat\bn\bx)\hat\bn+\sin |\bn| \hat\bn\wedge\bx & \text{for}\;\bn^2<0\\
\bx+\bn\wedge\bx & \text{for}\;\bn^2=0.
\end{cases}
\end{align}
The group of orientation and time orientation preserving isometries of (2+1)-dimensional Minkowski space is the (2+1)-dimensional Poincar\'e group $\text{Isom}(\MM^3)=ISO(2,1)^+_0$. It has a semidirect product structure $ISO(2,1)^+_0=SO^+_0(2,1)\ltimes \RR^3$. With the parametrisation
\begin{align}\label{poincpar} ISO^+_0(2,1)\ni (u,\ba)\qquad u\in PSU(1,1), \ba\in\RR^3,\end{align} its 
 group multiplication law takes the form
\begin{align}
\label{group multiplication}
(u_1,\ba_1)\cdot (u_2,\ba_2)=(u_1u_2, \ba_1+\Ad(u_1)\ba_2)\qquad\forall u_1,u_2\in PSU(1,1), \ba_1,\ba_2\in\RR^3,
\end{align}
and its action on Minkowski space is given by\begin{align}
\label{minkact}
(u,\ba)\bx=\Ad(u)\bx+\ba\qquad\forall \bx\in\RR^3.
\end{align}


\subsection{Vacuum spacetimes via the quotient construction}
\label{vacsum}

In the following, we consider Lorentzian (2+1)-gravity with vanishing cosmological constant and without matter. More specifically, we restrict attention to
maximal globally hyperbolic vacuum spacetimes with a complete Cauchy surface that are of topology 
$M\approx \RR^+\times S_g$, 
where $S_g$ is an orientable two-surface of genus $g\geq 2$.  Spacetimes of this type have been investigated extensively in mathematics and mathematical physics, and their properties are well-understood. They resemble the cosmological solutions of the Einstein equations in (3+1) dimensions such as Bianchi spacetimes.  In particular, it is shown in \cite{mess} that any spacetime of this type has a big bang singularity and is equipped with a cosmological time function. 

Due to the absence of local gravitational degrees of freedom in (2+1)-gravity, any vacuum solution of the (2+1)-dimensional Einstein equations with vanishing cosmological constant is flat and locally isometric to Minkowski space. However, the theory has a finite number  of global degrees of freedom  for  spacetimes of non-trivial topology. For the spacetimes considered in his paper, these global degrees of freedom are manifest in their description as quotients of regions in Minkowski space. 
It is shown in \cite{mess}, for a more recent and accessible discussion see \cite{bb,bg}, that these spacetimes are obtained as quotients of certain regions in (2+1)-dimensional Minkowski space by the action of cocompact Fuchsian groups.

 \subsubsection{The quotient construction for vacuum spacetimes}
\label{quotconstr}

 The first ingredient in the quotient construction is a  regular, 
 future complete domain  in (2+1)-dimensional Minkowski space. This is an open region $D\subset \MM^3$ with a distinguished set of points $D_0\subset \partial D$ consisting  of those points in $\partial D$ which admit spacelike support planes \cite{bb}. In the following we adopt the terminology of \cite{bb,bg} and refer to $D_0$ as the initial singularity of the domain $D\subset \MM^3$. Note, however, that this does not coincide exactly with the standard definition of a singularity via geodesic incompleteness  \cite{wald, hawkell}\footnote{As will become apparent in the next subsection, not only points in $D_0$ but all points in $\partial D$  are associated with  past-incomplete inextendible timelike geodesics. According to the standard definition, the initial singularity would therefore be the whole boundary $\partial D$ and not just  the subset $D_0\subset \partial D$.}.

\begin{figure}[h]
\centering
\includegraphics[scale=0.25]{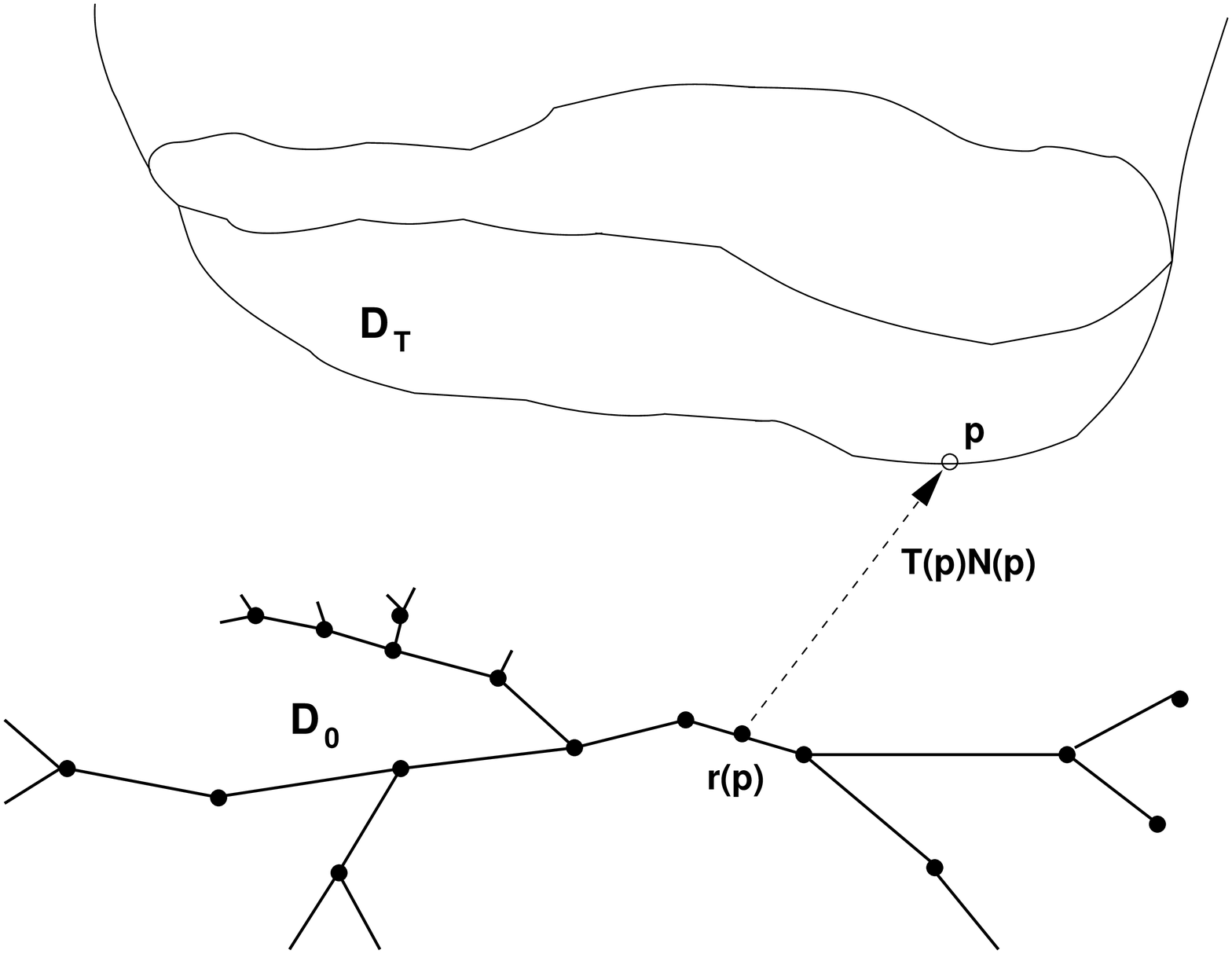}
\caption{\small{Parametrisation of points in the regular domain.}}
\label{dompar}
\end{figure}

It has been shown \cite{mess}, see also \cite{bb,bg}, that the regular domains in the quotient construction are equipped with a cosmological time function and foliated by surfaces $D_T$ of constant cosmological time $T$, i.~e.~of constant geodesic distance $T$ from the initial singularity $D_0$
\begin{align}
\label{domain}
D=\bigcup_{T\in \RR^+} D_T.
\end{align}
The domain is the future of the initial singularity $D_0$, and any point $\bp\in D$ can be parametrised uniquely as
\begin{align}
\label{pointparam}
\bp=T(\bp)\cdot N(\bp)+r(\bp)\qquad N(\bp)^2=-1, r(\bp)\in D_0,
\end{align}
where $T: D\rightarrow\RR^+$ is the cosmological time function,  $N:D\rightarrow H_1$ is called the Gauss map and takes values in the unit hyperboloid $H_1=\{\bx\in\mathbb M\,|\, \bx^2=-1\}\cong\hyp$ and $r:D\rightarrow D_0$ is  the retraction to the initial singularity $D_0$ as shown in Fig.~\ref{dompar}. 

The second ingredient in the construction is a cocompact Fuchsian group $\Gamma$ of genus $g$. This is a discrete subgroup of the proper orthochronous Lorentz group $ PSU(1,1)$ with $2g$ generators $v_{a_i}, v_{b_i}\in PSU(1,1)$, $i=1,\ldots,g$, and a single defining relation
\begin{align}
\label{fgroup}
\Gamma=\langle v_{a_1}, v_{b_1},..., v_{a_g}, v_{b_g} \in PSU(1,1)\;|\; [v_{b_g}, v_{a_g}^\inv]\cdots [v_{b_1}, v_{a_1}^\inv]=1\rangle,
\end{align}
where $[u, v]=u\cdot v\cdot u^\inv\cdot v^\inv$ denotes the group commutator.  
It can be shown that all non-unit elements of a cocompact Fuchsian group are hyperbolic, i.~e.~given as the exponential \eqref{su11rep} $v=\exp(n^aJ_a)$  with a spacelike vector $\bn$.

The cocompact Fuchsian group $\Gamma$ acts on the regular domain $D\subset\mathbb M$ via a group homomorphism  into the (2+1)-dimensional Poincar\'e group $ISO(2,1)^+_0=\text{Isom}(\mathbb M^3)$
\begin{align}
\label{grouphom}
h:\; &\Gamma\rightarrow ISO(2,1)^+_0,\quad
 v\mapsto h(v)=(\Ad(v),\ba(v)).\end{align}
It is shown in \cite{mess} that  this group homomorphism gives rise to a free and properly discontinuous action of  the cocompact Fuchsian group $\Gamma$  on the domain $D\subset \mathbb M$ which preserves each constant cosmological time surface $D_T$.

The quotient spacetime $M$ is obtained by taking the quotient of the regular domain $D$ with respect to this group action. As the latter preserves the constant cosmological time surfaces $D_T$ which foliate the domain, this amounts to identifying on each surface of constant cosmological time the points related by the action of $\Gamma$ via \eqref{grouphom}
\begin{align}
\label{spacetime}
M=\bigcup_{T\in\RR^+} M_T\qquad  M_T=D_T/ h(\Gamma).
\end{align}
In other words, two points $\bp,\bq\in D$ parametrised uniquely as in \eqref{pointparam} are identified if and only if  they satisfy $T(\bp)=T(\bq)$ and there exists an element $v\in\Gamma$ such that
\begin{align}
\label{explident}
N(\bq)=\Ad(v)N(\bp)\qquad r(\bq)=\Ad(v)r(\bp)+\ba(v).
\end{align}
The fact that the group action is free and properly discontinuous ensures that this quotient  is a three-dimensional manifold of topology $M\approx\RR^+\times S_g$ and that its fundamental group is isomorphic to the cocompact Fuchsian group   $\pi_1(M)\cong\pi_1(S_g)\cong\Gamma$. 
It inherits a Lorentzian metric induced by the restriction of the  Minkowski metric $\eta$ to the domain $D\subset \MM^3$. The fact that the group action preserves the surfaces of constant cosmological time implies that $M$ is equipped with a cosmological time function and foliated by spacelike surfaces $M_T=D_T/h(\Gamma)$ of constant cosmological time, i.~e.~of constant geodesic distance from  $M_0=D_0/h(\Gamma)$. As in the case of the domain, we adopt the terminology of \cite{bb,bg} and refer to $M_0$ as the initial singularity of $M$, although again this does not coincide exactly with the standard definition via geodesic incompleteness \cite{wald,hawkell}. 
 The metric on $M$ thus takes the form
$g=-dT^2+g_T$,
where $g_T$ is metric on constant cosmological time surface $M_T=D_T/h(\Gamma)$ induced by $\eta$.

The geodesics on the spatial surfaces $M_T$ are obtained as the quotients of geodesics on the constant cosmological time surfaces $D_T$ with respect to the group action \eqref{grouphom}. Closed geodesics based at a point $\bp\in M_T$ are thus in one-to-one correspondence with elements of the fundamental group $\pi_1(M)\cong\Gamma$.
They are given by $\Gamma$-equivalence classes of geodesics  $g:\RR\rightarrow D_T$ on the constant cosmological time surfaces $D_T$
that satisfy
\begin{align}
\exists v\in\Gamma:\quad g(t+t_v)=h(v)g(t)\qquad \forall t\in\RR.
\end{align}
The preimage of a closed geodesic $g:\RR\rightarrow M_T$ on the spatial surface $M_T$ is therefore a 
set of geodesics
$G_g=\{h(v)\tilde g\,|\, v\in\Gamma\}$, 
where $\tilde g:\RR\rightarrow D_T$ is a lift of $g$ to $D_T$ and $h(v)\tilde g$ is its image under the action of $v\in\Gamma$ via the group homomorphism \eqref{grouphom}.

Timelike future directed  geodesics on $M$ are given as $\Gamma$-equivalence classes of  timelike future directed geodesics in the domain. The latter can be parametrised as
\begin{align}
\label{geodtime}
g_{x,x_0}(t)=t\cdot \bx+\bx_0\qquad \bx^2=-1,\,x^0>0,\,\bx_0\in D,
\end{align}
where the parametrisation is unique up to a time shift 
\begin{align}
\label{timeshift}
t\mapsto t+t_0\qquad\bx_0\mapsto \bx_0-t_0\bx.
\end{align} The preimage of a timelike, future directed geodesic in the quotient spacetime $M$  is therefore a set of geodesics $G_{x,x_0}=\{h(v) g_{x,x_0}\,|\,v\in\Gamma\}$ where $g_{x,x_0}$ is a specific lift parametrised as in \eqref{geodtime} and $h(v)g_{x,x_0}$ is its image under the action \eqref{grouphom} of $v\in\Gamma$. Using the parametrisation \eqref{poincpar}, \eqref{minkact}, we find that these geodesics are of the form
\begin{align}
\label{geodim}
h(v) g_{x,x_0}(t)=t\cdot \Ad(v)\bx+\Ad(v)\bx_0+\ba(v)\qquad  \bx^2=-1, \;x^0>0, \;\bx_0\in D,\; v\in\Gamma.
\end{align}
Similarly, a future directed lightlike geodesic in $M$ corresponds to a set of geodesics $G_{x,x_0}=\{h(v)g_{x,x_0}\,|\, v\in\Gamma\}$  given as in \eqref{geodim} but with $\bx^2=0$. In this case, the parametrisation in terms of vectors $\bx, \bx_0\in\RR^3$ is unique up to a time shift \eqref{timeshift} and a rescaling $\bx\mapsto\alpha\bx$, $t\mapsto t/\alpha$ with $\alpha\in\RR^+$.

\subsubsection{Static spacetimes}

\begin{figure}[h]
\centering
\includegraphics[scale=0.5]{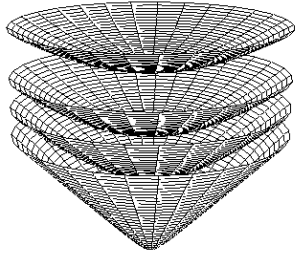}
\caption{\small{Foliation of the future lightcone by hyperboloids.}}
\label{lcfoli}
\end{figure}

We illustrate the general pattern of the construction by considering the simplest spacetimes obtained by it, the so-called static spacetimes\footnote{Strictly speaking, they are not static but {\em conformally static} \cite{wald}, but for simplicity we will refer to them as static spacetimes in the following. } associated to a cocompact Fuchsian group $\Gamma$.
In this case, the regular domain is the interior of a future lightcone based at a point $\bp\in\RR^3$
\begin{align}
\label{lightcone}
D^s= \{ \bx\in\MM^3\;|\: x^0-p^0>0, (\bx-\bp)^2<0\},
\end{align}
and the initial singularity is its basepoint  $D_0=\{\bp\}$. The 
foliation of the domain by surfaces of constant cosmological time is the usual foliation of the lightcone by hyperboloids   in Fig.~\ref{lcfoli}
\begin{align}
\label{hyps}
D^s_T=H_T+\bp=\{\bx\in\MM^3\;|\; (\bx-\bp)^2=-T^2,\; x^0-p^0>0\}.
\end{align}
As explained in the appendix,  the constant cosmological time surfaces $D_T^s$ with the metric induced by $\eta$ are isometric to the Poincar\'e disc model \eqref{unitdisc} of hyperbolic space $\hyp$ up to a rescaling of the metric with a factor $T^2$. 
Each constant cosmological time surface $D_T^s$ is therefore a copy of two-dimensional hyperbolic space   rescaled by the cosmological time $T$.  

The group homomorphism $h_s:\Gamma \rightarrow ISO_0^+(2,1)$ that acts on this static domain and preserves the constant cosmological time surfaces $D^s_T$ is given by
\begin{align}
\label{staticact}
h_s:\;v\mapsto (\Ad(v), \ba_s(v))=(\Ad(v), (1-\Ad(v))\bp).
\end{align}
From formula \eqref{minkact} for the action of the isometry group and the identification \eqref{hypdisc} between the hyperboloids and the Poincar\'e disc, it follows  that this group action agrees with the canonical action \eqref{isomact} of $\Gamma$ on the Poincar\'e disc. As explained in the appendix, this canonical action of $\Gamma$ induces a tessellation of hyperbolic space $\hyp$, and hence of the constant cosmological time surfaces $D_T^s$, by geodesic arc $4g$-gons, which are mapped into each other by the elements of $\Gamma$ as indicated in Fig.~\ref{fundpoly}.  

The spacetime $M$ is obtained by identifying on each hyperboloid the points related by the action of $\Gamma$ or, equivalently, by gluing the sides of each polygon pairwise as shown in Fig.~\ref{fundpoly}. It takes  the form
\begin{align}
\label{statmf}
M=\bigcup_{T\in\RR^+}  T \cdot \Sigma_g, \quad  \Sigma_g=\HH^2/\Gamma\qquad\quad g=-dT^2+T^2 g_{\Sigma_g}\end{align}
where  $g_{\Sigma_g}$ is the standard metric on the surface $\Sigma_g=\hyp/\Gamma$ induced by the metric on hyperbolic space $\hyp\cong H_1$.  The metric $g_T$ on the constant cosmological time surfaces $M_T$ therefore does not exhibit an interesting evolution with the cosmological time $T$.
 It is rescaled by an overall factor $T^2$ but stays proportional to the standard metric of the associated two-surface $\Sigma_g$. In the standard terminology \cite{wald}, the spacetimes are therefore conformally static, but for notational  simplicity we will refer to them as static in the following.

\subsubsection{Evolving spacetimes via grafting}

\begin{figure}[h]
\centering
\includegraphics[scale=0.3]{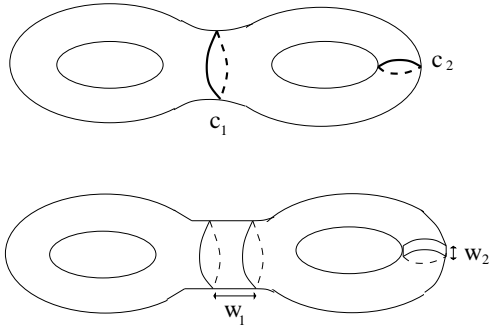}
\caption{\small{Grafting along a multicurve consisting of  two geodesics $c_1$, $c_2$ with weights $w_1$, $w_2$ on a genus 2 surface.}}
\label{graftsurf}
\end{figure}

It is shown in \cite{mess} that any maximally globally hyperbolic genus $g$ vacuum spacetime can be obtained from a static spacetime via the grafting construction. The ingredients in the grafting construction are a cocompact Fuchsian group $\Gamma$ and a measured geodesic lamination on the associated  surface $\Sigma_g=\hyp/\Gamma$. The measured geodesic lamination can be thought of as the limit of a sequence of weighted multicurves (for a precise definition of this limit see \cite{mess,bb}). These are sets of non-intersecting geodesics on $\Sigma_g$ with a positive number, the weight, associated to each geodesic as shown in Fig.~\ref{graftsurf}. In the following, we summarise the grafting construction for multicurves following the presentation in \cite{bb}.

Schematically, grafting acts on each constant cosmological time surface $M_T\cong T\cdot\Sigma_T$ in the static spacetime by inserting a strip along each geodesic in the multicurve on $\Sigma_g$ as shown in Fig.~\ref{graftsurf}. The construction is performed simultaneously on all surfaces $M_T$, and the widths of the strips are given by the weights of the associated geodesics.

\begin{figure}[h]
\centering
a)\includegraphics[scale=0.5]{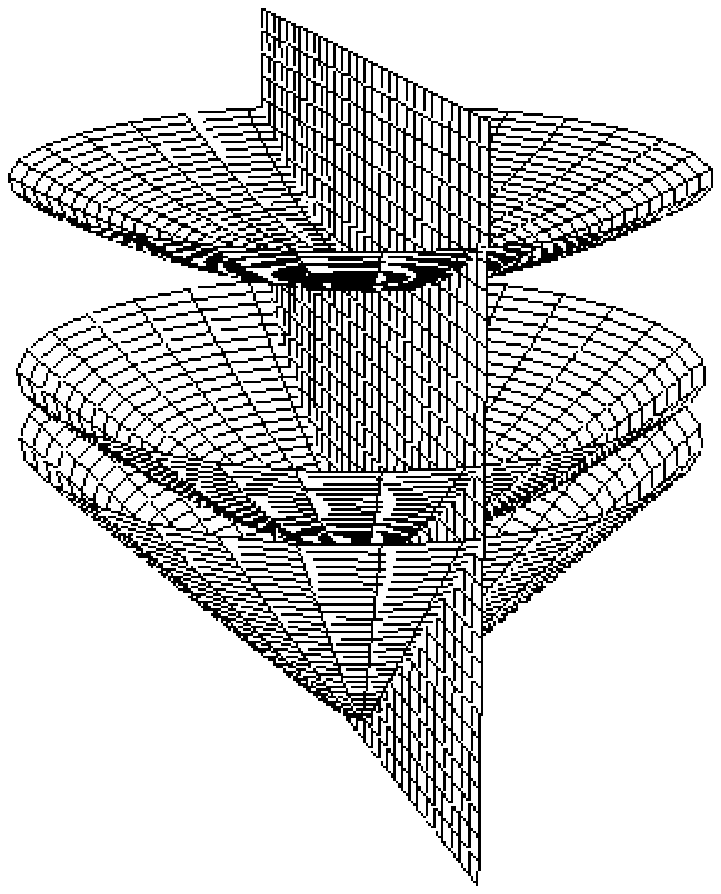}
b)\includegraphics[scale=0.5]{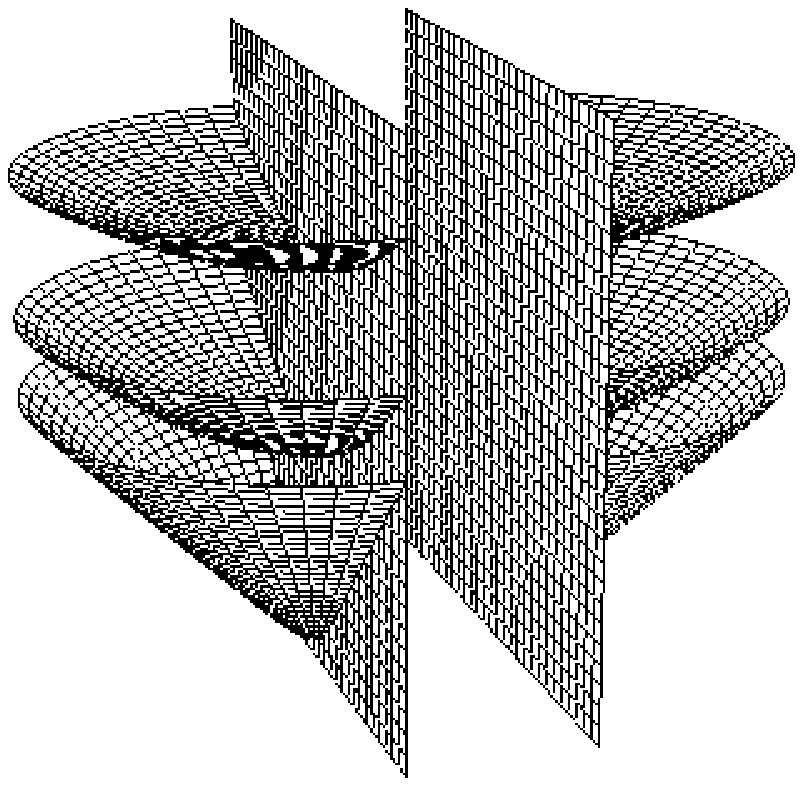}
c)\includegraphics[scale=0.5]{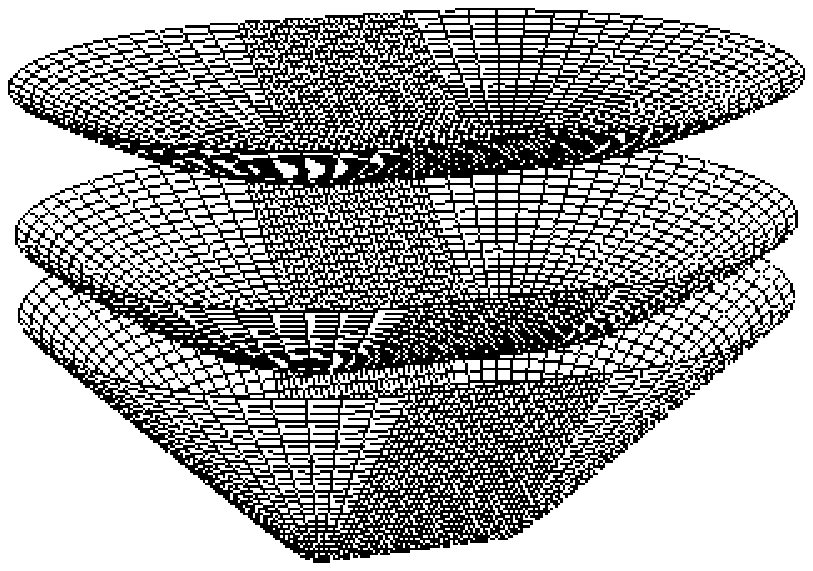}
\caption{\small{The grafting construction for a single geodesic in the regular domain $D_s$.}}
\label{grafting}
\end{figure}

The construction is performed in the universal cover, i.~e.~the regular domain $D_s\subset\MM^3$ and 
implemented via the following steps:
\begin{enumerate}
\item Lift the geodesics in the multicurve to a $\Gamma$-invariant set of geodesics on each of the hyperboloids  $D^s_T = T\cdot \hyp$ by selecting one lift for each geodesic and acting on it with the elements of  $\Gamma$. This yields a $\Gamma$-invariant weighted multicurve on each of the  constant cosmological time surfaces $D_T^s$ which foliate the interior of the future lightcone. The geodesics in this multicurve are given as intersections of planes through the tip of the lightcone with the hyperboloids $D_T^s$ as shown in Fig.~\ref{grafting} a.
\item Select a basepoint $\bq$ in the interior of the future lightcone $D_s$ outside of all the geodesics in the multicurve, i.~e.~outside the planes defining these geodesics.
\item Cut the lightcone $D_s$ along all of the planes corresponding to geodesics in the multicurve. Translate the pieces that do not contain the basepoint away from the basepoint in the direction of the normal vector of the associated plane and by a distance given by the weight of the associated geodesic as shown in Fig.~\ref{grafting} b.
\item Join the resulting pieces by straight lines connecting the two points that correspond to a given point on each geodesic in the multicurve as shown in Fig.~\ref{grafting} c.
\end{enumerate}

The result of the construction is a deformed domain $D$. Its initial singularity $D_0$ is no longer a point but a graph, more specifically, a $\Gamma$-invariant real simplicial tree. The surfaces $D_T$ of constant cosmological time $T$  which foliate the deformed domain are the images of the hyperboloids $D_T^s$ under the grafting construction. 
They are deformed hyperboloids with a strip glued in along each geodesic in the multicurve. 

The grafted spacetime is given as the quotient $M=D/h(\Gamma)$ of the deformed domain $D$ by a deformed action of the cocompact Fuchsian group $\Gamma$.  
This group action is defined in such a way that it identifies two points in the deformed domain if and only if the canonical action of  $\Gamma$ via \eqref{staticact} identifies the corresponding points in the static spacetime. (Points on the strips are identified if and only if the corresponding points on the geodesics in the multicurve are identified via \eqref{staticact} and  they have the same distance from the edge of the strips.) The associated group homomorphism $h:\Gamma\rightarrow ISO^+_0(2,1)$ therefore acquires a  translational component which takes into account the translations  in the grafting construction:
\begin{align}
\label{grafthol}
h(v)=(\Ad(v), (1-\Ad(v))\bp+\sum_{i=1}^k\lambda_i\hat\bn_i)=(1,\sum_{i=1}^k\lambda_i\hat\bn_i)\cdot h_s(v),
\end{align}
where the sum runs over all geodesic in the multicurve in $\hyp$ which intersect the geodesic segment from the basepoint $\bq\in\hyp$ to its image $\Ad(v)\bq\in\hyp$. The parameters $\lambda_i$ denote the weights of the geodesics and the vectors $\hat\bn_i$ the spacelike unit normal vectors of the associated planes oriented in such a way that $\bq\cdot \hat\bn_i<0$, $\Ad(v)\bq\cdot \hat\bn_i>0$.

\begin{figure}[h]
\centering
\includegraphics[scale=0.4]{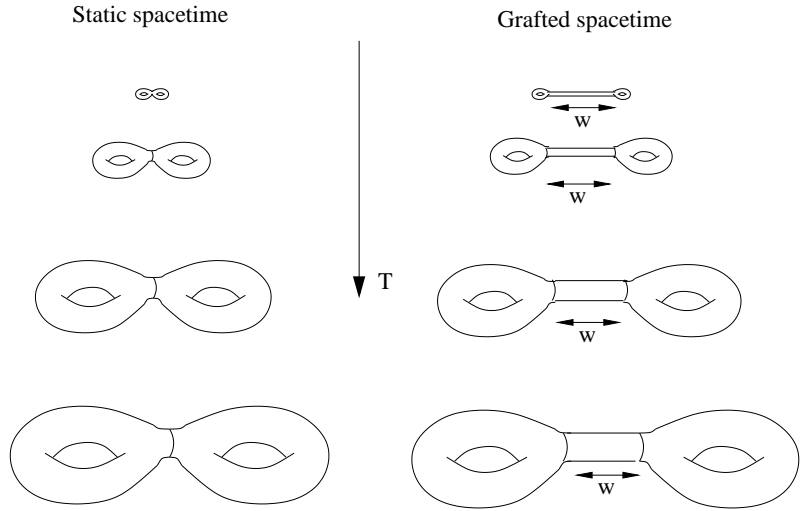}
\caption{\small{Illustration of the evolution of static and grafted spacetimes with the cosmological time.  While the hyperbolic part of the constant cosmological time surfaces $M_T$ is rescaled with the cosmological time $T$, the widths of the grafted strips stay constant, which yields a non-trivial evolution of the metric on the spatial surface.}}
\label{timevol}
\end{figure}

The resulting quotient spacetime $M=D/h(\Gamma)$ is no longer static. The metric $g_T$ on the surfaces $M_T=D_T/h(\Gamma)$ of constant cosmological time evolves with time as depicted in Fig.~\ref{timevol}. While the pieces of the constant cosmological time surfaces $M_T$ outside the strips are simply rescaled by the cosmological time $T$, the width of the strips is given by the weight of the grafting geodesics and stays constant. The metric $g_T$ thus evolves with the cosmological time $T$ and approaches the metric of the constant cosmological time surfaces in the associated static spacetime in the limit  $T\rightarrow\infty$


\subsection{Phase space and observables}
\label{subphsp}

As shown by Mess \cite{mess}, any maximal globally hyperbolic flat (2+1)-spacetime with a complete Cauchy surface of genus $g\geq 2$ can be obtained via the quotient construction summarised in Sect.~\ref{vacsum}. Moreover, the geometry of such spacetimes is determined uniquely by the choice of a cocompact Fuchsian group $\Gamma$ as in \eqref{fgroup} together with its action on Minkowski space via the group homomorphism \eqref{grouphom}. As the cocompact Fuchsian group $\Gamma$ is isomorphic to the fundamental group $\Gamma\cong\pi_1(M)$, this implies that every spacetime is characterised uniquely\footnote{This is in general not the case for  (2+1)-spacetimes with point particles. It has been shown by Matschull \cite{Matschull2}, that there exist examples of non-diffeomorphic spacetimes with identical holonomies. A mathematical discussion of this phenomenon of "holonomy failure" for the case of point particles on a sphere is given in \cite{bb}. For an investigation of the analogous phenomenon in (1+1)-dimensional gravity see \cite{SchS}.} by a group homomorphism  
\begin{align}
h:\; &\pi_1(M)\rightarrow ISO^+_0(2,1)\qquad \lambda\in\pi_1(M)\mapsto h(v_\lambda)=(\Ad(v_\lambda), \ba(v_\lambda))\in ISO^+_0(2,1). 
\end{align}
Equivalently, the spacetimes can be characterised by the values of this group homomorphism on a set of generators of the fundamental group $\pi_1(M)\cong\Gamma$, i.~e.~by the $ISO(2,1)^+_0$-valued holonomies\footnote{Note that these holonomies coincide with the ones obtained in the Chern-Simons formulation of the theory and defined as path ordered exponentials $H(c)=P\exp\int_0^1 A(c(t)) \dot c(t) dt$ where $A$ is the $\mathfrak{iso}(2,1)$-valued connection.} along a set of closed curves on the spatial surface $M_T$ that represent these generators. 

Two group homomorphisms $h_i:\pi_1(M)\rightarrow ISO^+_0(2,1)$, $i=1,2$, determine diffeomorphic spacetimes if and only if they are related by conjugation with a constant element of the (2+1)-dimensional Poincar\'e group $ISO^+_0(2,1)$. This  corresponds to a global Poincar\'e transformation acting simultaneously on the domain and on the holonomies according to
\begin{align}
\label{globpoinc}
&\by\mapsto \Ad(v_0)\by+\ba_0\quad h(v)\mapsto(\Ad(v_0),\ba_0)\cdot h(v)\cdot(\Ad(v_0),\ba_0)^\inv\qquad\forall \by\in D, v\in\Gamma.
\end{align}
Using the group multiplication law \eqref{group multiplication} and formula \eqref{minkact} for the Poincar\'e transformations, it follows directly that two points in the transformed domain $(\Ad(v_0),\by_0)D$ are related by the group action $h_2=(\Ad(v_0),\by_0)\cdot h_1\cdot (\Ad(v_0),\by_0)^\inv$ if and only if the corresponding points in the original domain $D$ are related by $h_1$.
As  Poincar\'e transformations are isometries of Minkowski space, the resulting  quotient spacetimes $M=D/h(\Gamma)$ are isometric.
This implies that the phase space of the theory is the identity component of the space of all group homomorphism from the fundamental group $\pi_1(M)$ into the (2+1)-dimensional Poincar\'e group $ISO^+_0(2,1)$ modulo conjugation with  $ISO^+_0(2,1)$
\begin{align}
\label{phspace}
\text{Hom}_0(\pi_1(M), ISO_0^+(2,1))/ISO_0^+(2,1).
\end{align}
The physical observables of the theory, which are functions on the phase space, are thus given as functions of these $2g$ $ISO^+_0(2,1)$-valued holonomies  that are invariant under simultaneous conjugation of their arguments with $ISO^+_0(2,1)$.

A specific set of such observables are the Wilson loop observables associated to closed curves in $M$. They were first investigated in \cite{RN, RN1,RN2,RN4,RN3,ahrss}  and have since played an important role in the classical description and the quantisation of the theory. They are
obtained by applying a conjugation invariant function $f: ISO^+_0(2,1)\rightarrow \RR$ to the curve's holonomy. Due to the absence of local degrees of freedom, they only depend on the  curve's homotopy equivalence class in $\pi_1(M)\cong\Gamma$ and can therefore be viewed as maps
\begin{align}
\label{wabstr}
W_f:\pi_1(M)\rightarrow\RR\qquad \lambda\mapsto f(h(v_\lambda)),
\end{align}
where $v_\lambda$ is the element of $\Gamma\cong\pi_1(M)$ associated to $\lambda$ and $h$ the group homomorphism \eqref{grouphom}.

For Lorentzian (2+1)-gravity with vanishing cosmological constant, each element $\lambda\in\pi_1(M)$ is associated with two canonical Wilson loop observables  which are the fundamental physical observables of the theory.  
It is shown in \cite{ich,ich3} that they generate via the Poisson bracket the two fundamental transformations that change the geometry of the (2+1)-spacetime, grafting and earthquake performed simultaneously on all surfaces of constant cosmological time. 
These canonical Wilson loop observables, in the following  referred to as "mass" $m_\lambda$ and "spin" $s_\lambda$  of  $\lambda\in\pi_1(M)$, are obtained by applying the functions $m,s:\,ISO^+_0(2,1)\rightarrow \RR$
\begin{align}
\label{massspindef}
m:\; (e^{n^bJ_b},\ba)\mapsto {|\bn|}\qquad s:\; (e^{n^bJ_b},\ba)\mapsto \ba\cdot \hat\bn
\end{align}
to the holonomy along $\lambda$. Note that they are closely related to the traces of the Poincar\'e-valued holonomies. Using the $\mathfrak{su}(1,1)$ representation \eqref{su11rep} one finds
\begin{align}
&\text{Tr}(e^{n^bJ_b})=2\cosh \left(\tfrac 1 2 m(e^{n^bJ_b}\!,\ba) \right)\quad
\text{Tr}(e^{n^bJ_b} \!\cdot\! a^cJ_c)=\sinh\left(\tfrac {1} 2  m(e^{n^bJ_b}\!,\ba)\right)\!\cdot\! s(e^{n^bJ_b}\!,\ba).
\end{align}
It has been shown that the mass and spin observables associated to all elements of the fundamental group $\pi_1(S_g)\cong\Gamma$ form a complete set of observables. Their values determine the spacetime uniquely and they parametrise the physical phase space \eqref{phspace} of the theory.


\section{Time, measurements and observers in (2+1)-gravity}

\label{obsmotiv}

\subsection{Time, measurements and observers}

\label{tmo}
After summarising the properties of maximally globally hyperbolic vacuum spacetimes in (2+1)-gravity with vanishing cosmological constant, we will now use these spacetimes to investigate the relation between spacetime geometry, the physical phase space of the theory and measurements by observers. This will yield concrete examples in which the conceptual issues surrounding time, measurements and the phase space of gravity are manifest and can be investigated. In particular, we will address the following  questions.
\begin{description}
\item[\bf 1. The physical phase space and measurements by observers] $\quad$

As explained in the previous section,  the phase space of (2+1)-gravity is finite-dimen\-sio\-nal and admits a simple parametrisation \eqref{phspace} in terms of holonomy variables or, equivalently, Wilson loop observables \eqref{wabstr}. These  variables  are the fundamental building blocks in most quantisation approaches.
 However, except for particularly simple cases such as the torus universe and certain point particle models, the physical interpretation of these holonomies and Wilson loops is currently not well understood. It is unclear how quantities that could be measured by an observer in the spacetime are given as functions of these holonomies  and, conversely, how the values of these observables could be determined by concrete measurements. This complicates the interpretation of the  theory and makes it difficult to extract interesting physics from the models. 

\item[\bf 2. The concept of observables] $\quad$

The issue of observables in constrained systems and, especially, gravity is subtle because the theory interlaces several  notions of  physical observables. The first is the concept of  
 physical observables as quantities that could be measured by an observer such as time, lengths, angles etc. The other is the notion of   observables as  functions on the { physical} phase space of the theory, i.~e.~the space of solutions of its equations of motion modulo gauge symmetries.  While these two notions coincide  for many physical systems, this is not the case for gravity. Functions on the physical phase space are by definition gauge and diffeomorphism invariant, while this is not the case for the usual quantities measured by observers such as lengths, areas or  time intervals. This issue gave rise to many discussions and lead to the development of the concepts of partial and complete observables and evolving constants of motion  by Rovelli \cite{rovobs1,rovtime1,rovtime2} and a formalism for the construction of complete (Dirac) observables by Dittrich \cite{bianca1, bianca2}.

\item[\bf 3. The role of time in the theory] $\quad$

As the Hamiltonian of general relativity is a constraint, there is no evolution of physical states in the phase space with respect to a time parameter. Physical states are labelled by the time-independent holonomy variables and Wilson loops. However, as explained in the previous section, the geometry of the constant cosmological time  surfaces evolves with respect to the cosmological time.
This implies in  particular that any realistic measurement by an observer will depend on time variables,  such as the cosmological time or the observer's eigentime. However, these time variables are not parameters that describe an evolution in phase space, but properties of the spacetime, i.~e.~the physical states themselves. This raises the question how such time variables enter the theory and manifest themselves in the relation between physical measurements and the gauge and diffeomorphism invariant observables. 
This issue is of special relevance to the subject of quantum gravity because it arises in many debates concerned with the structure  of time and space in a quantum theory of gravity.
\end{description}

The Lorentzian vacuum spacetimes considered in this paper appear as an ideal testing ground for the investigation of these questions. Due to the simplifications in (2+1) dimensions, their phase space can be parametrised explicitly in terms of gauge and diffeomorphism invariant observables and theory becomes amenable to quantisation.  Moreover, these spacetimes have a rich geometry with realistic physical features such as an initial singularity and expansion with the cosmological time. They also exhibit strong similarities with  the  Bianchi  models in (3+1) dimensions which are investigated extensively in cosmological applications of quantum gravity. They thus constitute viable  toy models for the (3+1)-dimensional case.

\subsection{The relation to gravitational lensing}
\label{gravlens}
The starting point for our investigation
is the question which quantities an observer in an empty (2+1)-spacetime could measure. Due to the absence of matter, all measurements of such an observer must be measurements of spacetime geometry itself. The description of spacetime via the quotient construction suggests quantities such as relative lengths of or angles between closed geodesics on surfaces of constant cosmological time. 
However, it turns out that expressing these quantities in terms of the holonomies which parametrise the phase space of the theory  is complicated for evolving spacetimes. It amounts to explicitly recovering the geodesic lamination  underlying  the grafting construction from the holonomies,    which is known to be difficult \cite{boncomm}.  

For this reason we pursue an alternative approach and consider an observer who determines the geometry of the spacetime by emitting lightrays.  Schematically, such an observer will notice that lightrays sent in certain directions return to him. He can determine the eigentime elapsed between the emission and reception of such returning lightrays, the  directions into which the light needs to be sent in order to return and the angles between them. Moreover, he can compare the relative frequency of the emitted and returning lightray and determine how all of these quantities evolve with respect to his eigentime at the emission of the lightray.   

The procedure is similar to  gravitational lensing (for an overview see \cite{wambs, perl}), which is used extensively in astrophysics and astronomy. In gravitational lensing, an observer probes the geometry of a spacetime region by observing multiple images of a light source behind it.
As in gravitational lensing, our observer makes use of multiple lightrays between two worldlines to determine the geometry of the spacetime. 

The situations differ insofar as in (3+1)-dimensional gravitational lensing the multiple images, aberration and frequency shifts of the lightrays  are due to the non-trivial gravitational field between the source and the observer. In (2+1)-gravity this gravitational field vanishes. Instead, the effect is caused by the nontrivial topology of the spacetime. In analogy to the (3+1)-dimensional case the procedure can therefore be viewed as a topological version of gravitational lensing. 
The other difference is that for reasons of simplicity we take the observer himself as a light source and consider lightrays that return to him, i.~e.~we consider measurements of the images of the lightsource as seen by the lightsource. However, the discussion in the next section should make it clear, how our  analysis can be generalised to external lightsources.


\section{Measurements  via returning lightrays}
\label{messect}

\subsection{Return time for lightrays emitted by an observer}

\label{rettime}

To investigate the notions outlined in the previous section, we now focus on measurements of an observer in free fall whose worldline is given by a future directed timelike geodesic in the spacetime $M$. As explained in Sect.~\ref{vacsum}, this corresponds to an infinite set of  future directed, timelike geodesics in the domain $D$, which are mapped into each other by the action of $\Gamma$ via \eqref{grouphom}. These geodesics can be  parametrised uniquely up to the time shift \eqref{timeshift} as  \begin{align}
\label{geodset}
&h(v)g_{x,x_0}(t)=t\!\cdot\!\Ad(v)\bx\!+\!\Ad(v)\bx_0\!+\!\ba(v)\quad\text{with}\quad \bx\in H_1, \bx_0\in D, t\in\RR^+,v\in\Gamma.
\end{align}
Lightrays emitted by the observer in $M$ at eigentime $t$ that return to him at time $t+\Delta t$ correspond to 
lightrays  in the domain $D\subset \MM^3$  that are emitted at the worldline $g_{x,x_0}=h(1)g_{x,x_0}$
at eigentime $t$ and are received at one of its images $h(v)g_{x,x_0}$ at time $t+\Delta t$. They  are thus in one-to-one correspondence with elements of the cocompact Fuchsian group $\Gamma\cong\pi_1(M)$. As explained in Sect.~\ref{quotconstr} the elements of $\Gamma\cong\pi_1(M)$ are also in one-to-tone correspondence with closed geodesics on each surface $M_T$ of constant cosmological time and, in particular, with geodesics on the static surfaces $M_T^s\cong T\cdot \Sigma_g$. We therefore have a one-to -one correspondence between returning lightrays and closed geodesics on the constant cosmological time surfaces $M_T$. 

The  interval $\Delta t$ of the observer's eigentime elapsed between the emission and reception of such a lightray is given by the condition
$\left(h(v)g_{x,x_0}(t+\Delta t)-g_{x,x_0}(t)\right)^2=0$. This yields a  quadratic  equation in $\Delta t$ with solutions 
\begin{align}
\label{sols}
\Delta t=\Ad(v)\bx\cdot\big(h(v)g_{x,x_0}(t)-g_{x,x_0}(t)\big)\pm \left| \Pi_{(\Ad(v)\bx)^\bot}\big(h(v)g_{x,x_0}(t)-g_{x,x_0}(t)\big)\right|,
\end{align}
where  $g_{x,x_0}(t)$ is given by \eqref{geodset} and $\Pi_{\bw^\bot}$ denotes the projection on $\bw^\bot$. In the following we focus on the plus sign in  \eqref{sols}  which characterises the future directed lightray.

To gain a better understanding of  this solution, we use the linear independence of 
 the vectors $\bx$, $\Ad(v)\bx$, $\bx\wedge\Ad(v)\bx$  for $v\in\Gamma\setminus\{1\}$  and characterise the initial translation vector $h(v)g_{x,x_0}(0)-g_{x,x_0}(0)$ in terms of  three parameters $\sigma_{v}, \tau_{v}, \nu_{v}\in\RR
$ for each $v\in\Gamma$
\begin{align}
\label{zparam2}
h(v)g_{x,x_0}(0)-g_{x,x_0}(0)=&\Ad(v)\bx_0\!-\!\bx_0\!+\!\ba(v)=\sigma_{v}(\Ad(v)\bx\!-\!\bx)\!+\!\tau_{v}\bx\!+\!\nu_{v}\,\bx\wedge\Ad(v)\bx. \end{align}
Moreover, we note that the scalar product $\bx\cdot \Ad(v)\bx$ is related to the geodesic distance \eqref{hypdistdef} of $\bx$ and $\Ad(v)\bx$ in the hyperboloid $H_1\cong \HH^2$ 
\begin{align}
\label{hypdistance}
&\bx\cdot \Ad(v)\bx=-\cosh \rho(\bx,\Ad(v)\bx)=:\cosh \rho_v.
\end{align}
This agrees with the length of the geodesic characterised by $\bx$ and $v\in\Gamma$ on the quotient surface $ \Sigma_g=\hyp/\Gamma=M_1^s$, i.~e.~on the surface of cosmological time $T=1$ of the corresponding static spacetime. 

Inserting \eqref{geodset}, \eqref{zparam2} and \eqref{hypdistance} into \eqref{sols}, 
we  find that the interval $\Delta t$ of eigentime elapsed between the emission and reception of the returning lightray associated to $v\in \Gamma$ is given by
\begin{align}
\label{nicesol}
\Delta t(t,v,\bx,\bx_0)=(\cosh\rho_v-1) (t+\sigma_{v})-\tau_{v}+ \sinh\rho_v\sqrt{(t+\sigma_{v})^2+\nu_{v}^2},
\end{align}
where $t$ is the observer's eigentime at the emission of the lightray, the variables
 $\sigma_v,\tau_v,\nu_v$ are defined by equation \eqref{zparam2} and $\rho_v$ by \eqref{hypdistance}. All of the parameters $\rho_v$, $\sigma_v$, $\tau_v$, $\nu_v $ in \eqref{nicesol} are  given as functions of the vectors $\bx\in H_1$, $\bx_0\in D$, which characterise the observer, and of the holonomies $h(v)=(\Ad(v),\ba(v))$, $v\in\Gamma$, which characterise the spacetime. 

{\bf Gauge invariance}

Expression \eqref{nicesol}  is invariant under the shift \eqref{timeshift} of the origin of the observer's eigentime, which reflects the redundancy in the parametrisation \eqref{geodset} of his worldline.
Combining equation \eqref{timeshift} and \eqref{zparam2}, one finds that this timeshift manifests itself as a transformation $t\rightarrow t+t_0$, $\sigma_v\rightarrow \sigma_v-t_0$ $\forall v\in\Gamma$, while the parameters $\rho_v,\tau_v,\nu_v$ are unaffected. This leaves  \eqref{nicesol} invariant. 

Moreover, equation \eqref{nicesol} is invariant under  global Poincar\'e transformations \eqref{globpoinc} acting simultaneously on the points in the domain and on the holonomies. Under such transformations, the vectors $\bx\in H_1$, $\bx_0\in D$  characterising the observer's worldline transform according to \eqref{globpoinc}, while all holonomies $h(v)$, $v\in\Gamma$ are conjugated. Using equations \eqref{zparam2}, \eqref{hypdistance}, one finds that the parameters $\rho_v$, $\sigma_v$, $\tau_v$, $\nu_v$ in \eqref{nicesol} are invariant under such  transformations for all $v\in\Gamma$ and hence equation \eqref{nicesol} is preserved.
 
In particular, this implies that the time intervals are independent of the choice of the lift \eqref{geodset} of the observer's  worldline in the domain $D$.  Considering instead the situation in which the lightray is emitted at eigentime $t$ at a geodesic $h(u)g_{x,x_0}$, $u\in\Gamma$ and received at time $t+\Delta t$ at $h(uv u^\inv)g_{x,x_0}$ corresponds to a Poincar\'e transformation \eqref{globpoinc} with $(v_0,\ba_0)=h(u)$. 
In that sense, the intervals of eigentime elapsed  between the emission and reception of a returning lightray are  diffeomorphism invariant quantities that characterise the spacetime. 
 
{\bf Observers in static spacetimes}

To gain a better understanding of formula \eqref{nicesol}, we consider the static spacetimes associated to a cocompact Fuchsian group $\Gamma$.
As discussed in Sect.~\ref{vacsum}, the group homomorphism \eqref{grouphom} then takes the form \eqref{staticact}  and the translational components of the holonomies are thus characterised by the condition $\ba(v)=(1-\Ad(v))\bp$ for a fixed vector $\bp\in\RR^3$ and all $v\in\Gamma$.  Observers whose worldline extends to the initial singularity and whose time origin coincides with the big bang are characterised by the condition $\bx_0=\bp$, where $\bx_0$ is the initial position of the observer  in \eqref{geodset}.
Inserting these conditions into \eqref{zparam2}, one finds that the parameters $\sigma_v,\tau_v,\nu_v$ vanish for all $v\in\Gamma$. Hence, for such an observer the eigentime elapsed between the emission and reception of a returning lightray  is a linear function of the eigentime at emission for {all}  returning lightrays 
\begin{align}
\label{stattime}
\Delta t_s(t,v,\bx)=t \cdot (e^{\rho_v}-1)\qquad \forall v\in\Gamma.
\end{align}
Static spacetime are therefore characterised by a linear relationship between the time interval $\Delta t$ and the eigentime $t$ at the emission of the lightray for  observers whose worldline extends to the initial singularity. The proportionality coefficient is given by the length $\rho_v$ of the closed geodesic associated with $v\in\Gamma$ on the static  surface $M_1^s=\hyp/\Gamma$.

{\bf Observers in evolving spacetimes}

\begin{figure}[h]
\centering
a)$\quad$\includegraphics[scale=0.3]{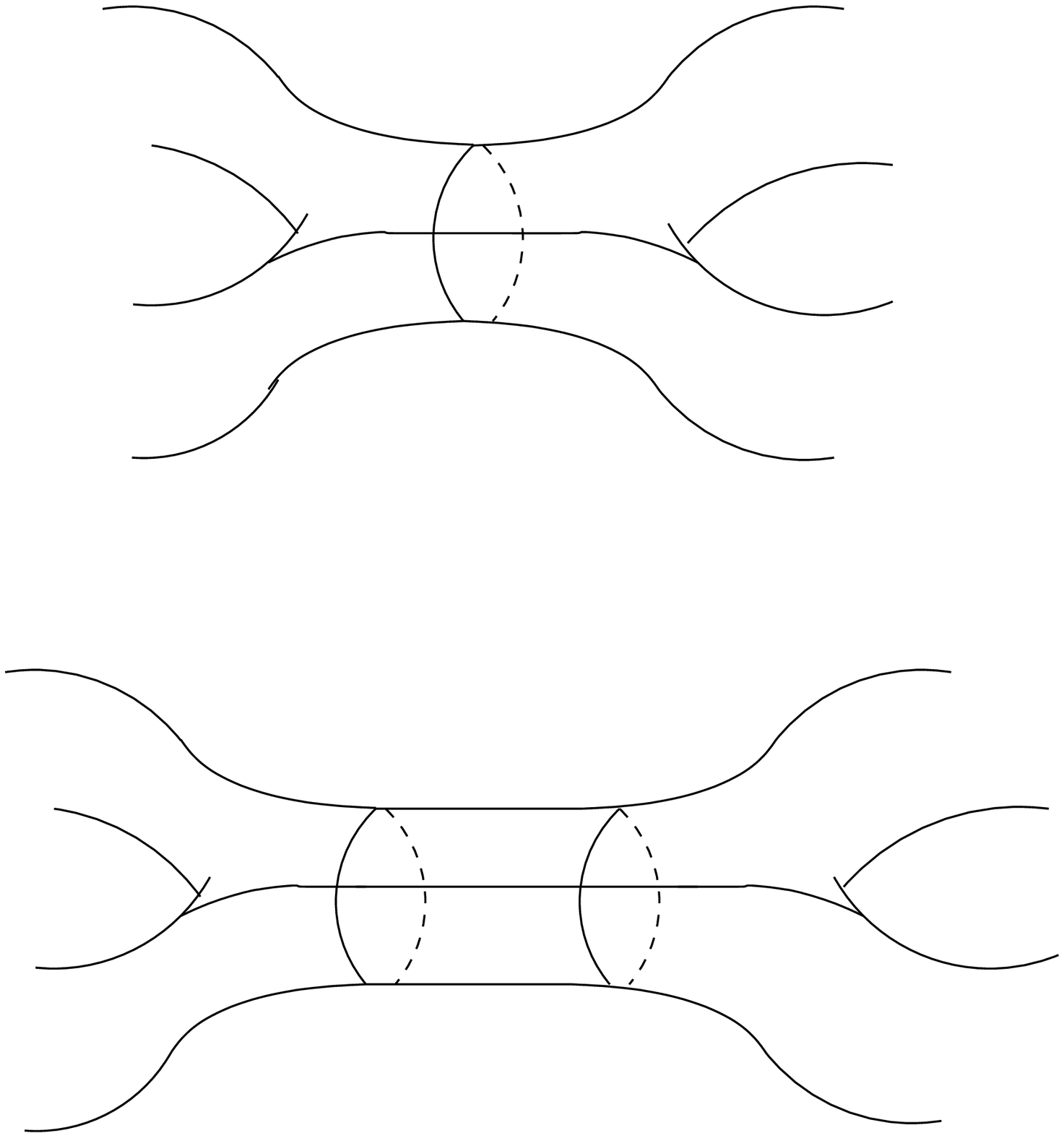} $\qquad\qquad\qquad$ b)$\quad$
\includegraphics[scale=0.3]{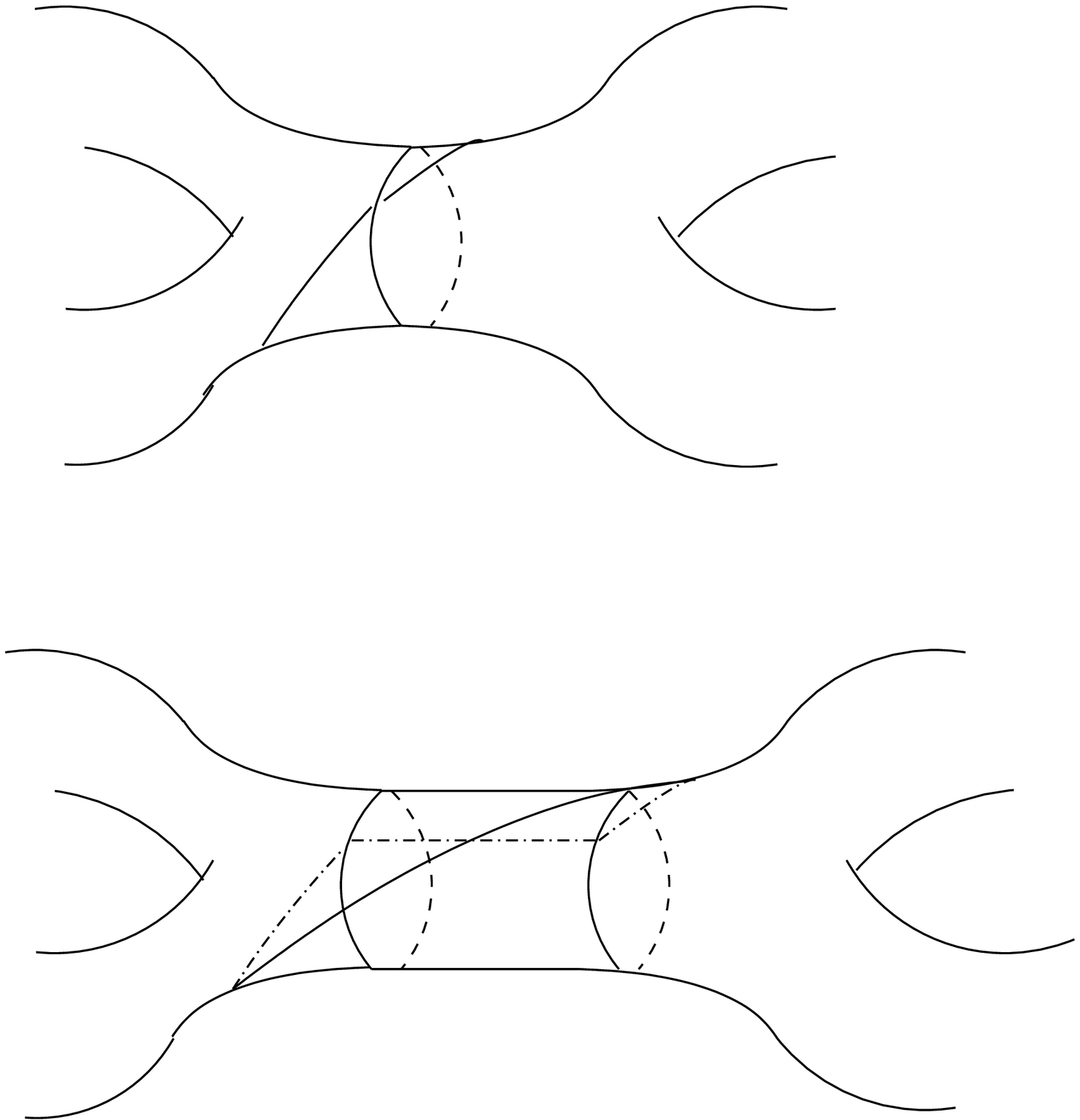}
\caption{\small{Deflection of geodesics at the grafted strips. The upper pictures show geodesics on the associated static surface, the lower pictures the corresponding geodesics on the grafted surface. Geodesics  which cross the grafted strip orthogonally (a) are not deflected, while all other geodesics (b) change their direction. }}
\label{geoddefl}
\end{figure}

For a general observer in an evolving spacetime, the eigentime $\Delta t$ elapsed between the emission and reception of a returning lightray is linear in the emission time $t$  if and only if the parameter $\nu_v$ in \eqref{nicesol} vanishes, i.~e.~if
translation vector \eqref{zparam2} between the two timelike geodesics $g_{x,x_0}$, $h(v)g_{x,x_0}$ in the domain lies in the plane spanned by $\bx$ and $\Ad(v)\bx$. The discussion in Sect.~\ref{vacsum} implies that this is the case if the corresponding geodesic on a spatial surface $M_T$ either does not cross the strips glued in via the grafting construction or crosses these strips orthogonally as depicted in Fig.~\ref{geoddefl} a. 
In this case, the direction of the geodesic does not change and its length increases by a constant contribution given by the width of the strip, which does not depend on the cosmological time. Hence, the eigentime elapsed between the emission and reception of the corresponding returning lightray  is modified with respect to the associated static spacetime by a constant contribution independent of the emission time $t$.

In contrast,  if the light is sent along a geodesic which crosses a strip as shown in Fig.~\ref{geoddefl} b, the direction of this geodesic is changed with respect to the corresponding geodesic on the static surface. This deflection of the geodesic depends on the cosmological time $T$, since the width of the strip is constant, while the rest of the surface grows linearly with $T$ as shown in Fig.~\ref{timevol}. The length of this geodesic therefore does not change linearly with the cosmological time. The time  \eqref{nicesol} elapsed between the emission and reception of the corresponding returning lightray therefore acquires a non-linear dependence on the emission time $t$.

However, for all observers in evolving spacetimes, formula  \eqref{nicesol} for the time intervals between the emission and reception of a returning lightray  approaches the expression for the associated static spacetime in the limit $t\rightarrow\infty$
\begin{align}
\label{limtime}
\lim_{t\rightarrow\infty} \frac{\Delta t (t, v, \bx,\bx_0)}{t}=e^{\rho_v}-1=\frac{\Delta t_s(t,v,\bx)}{t}.
\end{align}
The geometrical features which characterise the spacetime near the initial singularity and the parameters encoding the observer's initial position thus become redundant in this limit.

{\bf Relation to the Wilson loop observables}

While the dependence of  equation \eqref{nicesol} on the initial position of the observer vanishes in the limit $t\rightarrow\infty$, the motion of the observer in relation to the spacetime, i.~e.~his reference frame specified by the vector $\bx$ in \eqref{geodset} enters formulas \eqref{stattime}, \eqref{limtime} through the geodesic distance $\rho_v$. With the parametrisation $v=\exp(n_v^aJ_a)\in\Gamma$ via the exponential map \eqref{su11rep}  one obtains from equation \eqref{adrep}
\begin{align}
\label{hypdistad}
\cosh \rho_v=\cosh\rho(\bx,\Ad(v)\bx)=(\cosh|\bn_v|-1)(1+(\bx\bn_v)^2)+1.
\end{align}
For fixed $v\in\Gamma$, the time elapsed between the emission and reception of a returning lightray in the limit $t\rightarrow\infty$ is thus minimal for observers characterised by the condition $\bx\cdot\bn_v=0$. These are the observers whose  momentum vector  becomes parallel to the worldline of a point on the axis of the group element $v\in\Gamma$ in the limit $t\rightarrow\infty$.
For such observers, the parameters $\rho_v$, $\nu_v$ are given by the Wilson loop observables \eqref{massspindef}
\begin{align}
\label{msaxis}
&\rho_v=m_v & &\nu_v=s_v/\sinh m_v,
\end{align}
and the general expression \eqref{nicesol} for  time  intervals takes the form
\begin{align}
\label{axobs}
\Delta t=(\cosh m_v-1)(t+\sigma_v)-\tau_v+\sinh m_v\sqrt{(t+\sigma_v)^2+ s_v^2/\sinh^2 m_v}.
\end{align}
The mass observable $m_v$ therefore characterises the time elapsed between the emission and reception of the lightray   in the limit $t\rightarrow\infty$, while the spin observable $s_v$ characterises the non-linearity of the function $\Delta t(t)$ near the initial singularity. Note also that \eqref{adjact} and \eqref{zparam2}  imply that for any observer on the axis of $v\in\Gamma$ the spin observable $s_v$ does not depend on his initial position $\bx_0$, which only affects the parameters $\sigma_v,\tau_v$ in \eqref{axobs}.

We thus find that expression \eqref{nicesol} for the time elapsed between the emission and reception of a returning lightray provides a direct and physically  intuitive interpretation for the Wilson loop observables. They characterise the time elapsed between the emission and reception of a returning lightray  as measured by observers whose momentum vector becomes parallel to the worldline of a point on the associated geodesic in the limit $t\rightarrow\infty$.

\subsection{Angles and directions}

\label{dirangle}

{\bf Directions for returning lightrays}

To deepen our understanding of the relation between spacetime geometry  and the physical observables of the theory, we now consider the directions into which an observer needs to emit light in order to obtain returning lightrays and determine the angles between them.

As in the previous subsection, we consider an observer who emits a lightray at eigentime $t$ which returns to him at time $t+\Delta t$. 
The direction  into which this lightray is emitted in the momentum rest frame of the observer is characterised by a spacelike unit vector $\hat\bp_v(t)$ which is given as the projection of the vector characterising the lightray on the orthogonal complement of the observer's momentum vector
\begin{align}
\label{projvecdef}
\hat\bp_{v}(t)=&\frac{\Pi_{\bx^\bot}\left(h(v)g_{x,x_0}(t+\Delta t)-g_{x,x_0}(t)\right)}{|\Pi_{\bx^\bot}\left(h(v)g_{x,x_0}(t+\Delta t)-g_{x,x_0}(t)\right)|}.
\end{align}
Using formulas \eqref{tangproj}, \eqref{nicesol} and \eqref{zparam2}, we obtain
\begin{align}
\label{direval}
\hat\bp_{v}(t)=\cosh\varphi_v (t) \cdot \frac{\Pi_{\bx^\bot}(\Ad(v)\bx)}{|\Pi_{\bx^\bot}(\Ad(v)\bx)|}+\sin\varphi_v(t) \cdot \frac{\bx\wedge\Ad(v)\bx}{|\bx\wedge\Ad(v)\bx|},
\end{align}
where $\rho_v$ is given by \eqref{hypdistance}, $\sigma_v$, $\nu_v$ by \eqref{zparam2} and \begin{align}
\label{angledefl}
&\tan \varphi_v(t)=\frac{\nu_v}{f_v(t)}\qquad f_{v}(t)=(t+\sigma_{v})(\cosh\rho_{v}+1)+\sinh\rho_{v}\sqrt{(t+\sigma_{v})^2+\nu_{v}^2}.
\end{align}
As expected, the angles \eqref{angledefl}  are invariant under shifts \eqref{timeshift} of the observer's time origin, which do not affect the parameters $\rho_v$, $\tau_v$, $\nu_v$  and transform the variables $t$, $\sigma_v$ according to $t\rightarrow t+t_0$, $\sigma_v\rightarrow \sigma_v-t_0$ for all $v\in\Gamma$.  Moreover, they are invariant under 
 the global Poincar\'e transformations \eqref{globpoinc} acting simultaneously on the 
 parameters $\bx\in H_1,\bx_0\in D$, which characterise the
 observer's wordline, and on all holonomies . As equations \eqref{globpoinc}, \eqref{zparam2}, \eqref{hypdistance} imply that the parameters $\rho_v$, $\sigma_v$ and $\nu_v$ are invariant under such transformations, the deflection angles \eqref{angledefl} are preserved  and  the direction vectors \eqref{direval} transform covariantly.

{\bf Geometrical interpretation}

As discussed in the previous subsection,  an observer in a static spacetime whose worldline extends to the initial singularity is characterised by the condition  $\sigma_v=\tau_v=\nu_v=0$ for all $v\in\Gamma$. Hence, for such an observer  the angle \eqref{angledefl} vanishes for all values of the emission time $t$. 
The direction in which the lightray needs to be sent in order to return therefore is constant and coincides with the direction of the associated closed, spacelike geodesic.

For a general observer in an evolving spacetime, this direction is approached in the limit where the eigentime tends to infinity  $\lim_{t\rightarrow\infty } \varphi_v(t)=0$.
 Hence, a general observer finds that the directions into which light needs to be sent in order to return depend on the emission time but become constant in the limit $t\rightarrow\infty$, where they approach the ones for the associated static spacetime. 
 
The time dependence of the emission angle \eqref{angledefl}  is due to the the deflection of geodesics on the constant cosmological time surfaces $M_T$ at the grafted strips,  which is depicted in Fig.~\ref{geoddefl}. As explained  in the previous subsection, the direction of the geodesic changes with respect to the associated geodesic on the static surface if and only if it crosses strips non-orthogonally as depicted in Fig.~\ref{geoddefl} b. This deflection vanishes in the limit $T\rightarrow\infty$, as the width of the strips is constant but the rest of the surface is rescaled with a factor $T$.
If the geodesic associated with $v\in\Gamma$ does not cross any strips or crosses them orthogonally as in Fig.~\ref{geoddefl} a, the parameter $\nu_v$ in \eqref{angledefl} vanishes and its direction coincides with the one of the associated geodesic in the static surface for all values of the emission time $t$.

{\bf Angles between returning lightrays}

Although the deflection angles \eqref{angledefl} can in principle be measured by an observer, the comparison of directions with the directions approached in the limit $t\rightarrow\infty$ is impractical. We therefore consider the angle $\Phi_{v,w}(t)$ between two returning lightrays, both emitted at time $t$ and associated with group elements $v,w\in\Gamma$.
Using equations \eqref{direval} and  \eqref{angledefl}, we find that this angle is given as a sum of two contributions
\begin{align}
\label{angledef}
&\cos \Phi_{v,w}(t)=\hat\bp_{v}(t)\cdot \hat\bp_{w}(t)=\Phi_{v,w}(\infty)+\Psi_{v,w}(t).
\end{align}
The angle $\Phi_{v,w}(\infty)$, obtained in the limit $t\rightarrow\infty$, coincides with the one measured by an observer in the associated static spacetime whose worldline extends to the initial singularity
\begin{align}
\label{anglecomp1}
\Phi_{v,w}(\infty)=\lim_{t\rightarrow\infty} \Phi_{v,w}(t)=\arctan\left(\frac{\bx\cdot(\Ad(v)\bx\wedge\Ad(w)\bx)}{\Pi_{\bx^\bot}(\Ad(v)\bx)\cdot \Pi_{\bx^\bot}(\Ad(w)\bx)}\right).
\end{align}
The angle $\Psi_{v,w}$ is time-dependent and vanishes in the limit $t\rightarrow\infty$. It is given by
\begin{align}
\label{anglecomp2}
\Psi_{v,w}(t)=\arctan\left(\frac{\nu_{w}f_{v}(t)-\nu_{v}f_{w}(t)}{\nu_{v}\nu_{w}+f_{v}(t)f_{w}(t)}\right),
\end{align}
where the functions $f_v,f_w$ are defined as in \eqref{angledefl},  the parameters $\rho_v$, $\rho_w$ are given by \eqref{hypdistance} and  $\sigma_v,\sigma_w,\nu_v,\nu_w$  by \eqref{zparam2}. 
This angle describes the deflection of the two associated geodesics on the constant cosmological time surfaces $M_T$ at the strips glued in via the grafting construction and therefore varies non-trivially with 
the emission time $t$.

{\bf Relation to the Wilson loop observables}

For an observer whose momentum vector becomes parallel to the worldline of a point on the axis of the group element $v\in\Gamma$ in the limit $t\rightarrow\infty$, the parameters $\rho_v$, $\nu_v$ are given by \eqref{msaxis}. Expression \eqref{angledefl} for the deflection angle and formulas \eqref{angledef} and \eqref{anglecomp1}, \eqref{anglecomp2} for the angles between the direction of returning lightrays are therefore again given as functions of the two fundamental Wilson loop observables associated with $v\in\Gamma$. The mass observables $m_v$ characterise the measurements of the directions and angles for $t\rightarrow\infty$. The spin observables $s_v$  determine these measurements near the initial singularity of the spacetime. 

As in the case of the time elapsed between the emission and reception of a returning lightray, we thus find that the quantities measured by an observer - the directions into which the light needs to be emitted to return and the angles between those directions -  are given by gauge invariant functions on the phase space and  directly related  to the physical observables of the theory. The mass and spin observables \eqref{massspindef} arise naturally in the measurements of an observer whose momentum vector becomes parallel to the wordline of a point on the axis of $v\in\Gamma$ in the limit $t\rightarrow\infty$.

\subsection{Redshift}

\label{redshiftsec}

It is well-known that expanding cosmological solutions of the Einstein equations in (3+1) dimensions are associated with a redshift which serves as the basis for many cosmological measurements.  As the (2+1)-dimensional vacuum spacetimes considered in this paper have similar geometrical properties and also expand with the cosmological time, it is natural to ask if such a redshift is also present  in these spacetimes.

To determine if redshifts occur and to derive an explicit expression in terms of the physical  observables, we again focus an observer in free fall in the spacetime who emits a lightray at eigentime $t$ which returns to him at eigentime $t+\Delta t$.  As explained in the previous subsections, this corresponds to a lightray in the regular domain $D\subset \MM^3$ which is emitted at a future directed timelike geodesic at time $t$ and received at one of the geodesic's images under the action of $\Gamma$ at time $t+\Delta t$.  

Using this description in terms of geodesics in the regular domain,  the relative frequencies of the emitted and returning lightray  can be calculated straightforwardly using the relativistic Doppler effect.  
For this, we consider  two observers in Minkowski space with worldlines
$g_{i}(t)=\bz_i t+\bp_i$, $\bz_i^2=-1$, $\bp_i\in\RR^3$, $i=1,2$.  The relative frequencies of a lightray characterised by a vector $\bv\in\RR^3$, $\bv^2=0$, in the reference frames of these observers 
are
\begin{align}
\label{frequshift}
\frac{f_2}{f_1}=\frac{\bz_2\cdot \bv}{\bz_1\cdot \bv}.
\end{align}
In our situation, the two observers are replaced by two future oriented timelike geodesics in the regular domain, a  geodesic $g_{x,x_0}$, which lifts the worldline of the observer, and its image $h(v)g_{x,x_0}$ under the action of an element $v\in\Gamma$, both parametrised as in \eqref{geodset}.  A lightray emitted by the observer at time $t$ and returning to him at time $t+\Delta t$ is characterised by the lightlike vector $\bv=h(v)g_{x,x_0}(t+\Delta t)-g_{x,x_0}(t)$ with $\Delta t$ given by \eqref{nicesol}.
Hence, we have $\bz_1=\bx$, $\bz_2=\Ad(v)\bx$, and using   the parametrisation \eqref{zparam2} we obtain
\begin{align}
\bv=&(t+\sigma_v)(\Ad(v)\bx-\bx)+(\Delta t+\tau_v)\Ad(v)\bx+\nu_v\bx\wedge\Ad(v)\bx.
\end{align} 
Inserting this expression together with \eqref{nicesol} into \eqref{frequshift} and denoting by $f_e$ and $f_r$, respectively, the frequencies of the emitted and the returning lightray as measured by the observer, we obtain an expression for the relative shift in frequency
\begin{align}
\label{redshift}
\frac{f_r}{f_e}(t)=\frac{\sqrt{(t+\sigma_v)^2+\nu_v^2}}{\cosh\rho_v\sqrt{(t+\sigma_v)^2+\nu_v^2} +\sinh\rho_v(t+\sigma_v)},
\end{align}
where $\rho_v$  is given by \eqref{hypdistance} and the parameters $\sigma_v$, $\nu_v$ by \eqref{zparam2}. Note that this frequency shift is again a physical observable in the sense that it is invariant under the time shift  \eqref{timeshift} and under the global Poincar\'e transformations \eqref{globpoinc},  which act simultaneously on the observer's reference frame and on the holonomies.

{\bf Observers in static spacetimes}

For an observer in a static spacetime whose worldline extends to the initial singularity,  we have $\nu_v=0$ for all $v\in\Gamma$, and the frequency shift \eqref{redshift} takes the form
\begin{align}
f^s_r/f^s_e=e^{-\rho_v}.
\end{align}
Such an observer therefore measures a constant redshift which does not depend on the emission time. This redshift increases exponentially  with  the length of the associated geodesic  on the static constant cosmological time surface  $M^s_1\cong\hyp/\Gamma$. 

{\bf Observers in evolving spacetimes}

For a general observer in an evolving spacetime the function $f_r/f_e(t)$ in \eqref{redshift}  decreases monotonically
and approaches  the value for the corresponding static spacetime for $t\rightarrow\infty$ \begin{align}
\lim_{t\rightarrow \infty} f_r/f_e(t)=e^{-\rho_v}=f_r^s/f_e^s.
\end{align}
Hence, the redshift is maximal in the limit  $t\rightarrow\infty$ and minimal near the initial singularity. This raises the question if blueshifts can occur for observers near the initial singularity of an evolving spacetime. To demonstrate that this is not the case,  we consider an observer whose worldline extends to the initial singularity of an evolving spacetime and for whom the big bang coincides with the time origin $t=0$. 
As the function $f_r/f_e(t)$ decreases monotonically, it is maximal for $t=0$, where it takes the value
\begin{align}
f_r/f_e(0)=\frac{\sqrt{\sigma_v^2+\nu_v^2}}{\cosh\rho_v\sqrt{\sigma_v^2+\nu_v^2}+\sigma_v\sinh\rho_v}.
\end{align}
Hence, a blueshift at $t=0$ would occur if and only if
\begin{align}
\label{sigcond}
\sigma_v<-\tanh(\tfrac 1 2 \rho_v)\sqrt{\sigma_v^2+\nu_v^2}<0.
\end{align}

However, the grafting construction of evolving spacetimes summarised in Sect.~\ref{vacsum} implies \begin{align}
\label{grsig}
h(v)g_{x,x_0}(0)-g_{x,x_0}(0)=\sigma_v(\Ad(v)\bx-\bx)+\tau_v\Ad(v)\bx+\nu_v\bx\wedge\Ad(v)\bx=\sum_{i=1}^k \lambda_i \hat\bn_i,
\end{align}
where $\hat\bn_i$ are the spacelike unit normal vectors of the grafting geodesics in $\hyp$ which intersect the geodesic  segment from $\bx$ to $\Ad(v)\bx$ and the parameters  $\lambda_i\in\RR^+$ the associated weights. As explained in the paragraph after equation \eqref{grafthol}, the grafting construction requires that the unit normal vectors $\hat\bn_i$ of the grafting geodesics are oriented in such a way that
\begin{align}
\label{grcond}
&\Ad(v)\bx\cdot\hat\bn_i\geq0\qquad \bx\cdot\hat\bn_i\leq0\qquad\forall i=1,\ldots,k.
\end{align}
Inserting this condition into \eqref{grsig} and using the identity \eqref{hypdistance} for $\rho_v$, we find
\begin{align}
&\sigma_v(1\!-\!\cosh\rho_v)\!-\!\tau_v\cosh\rho_v=\sum_{i=1}^k \lambda_i \bx\cdot\hat\bn_i\leq 0\quad
\sigma_v(\cosh\rho_v\!-\!1)\!-\!\tau_v=\sum_{i=1}^k \lambda_i \Ad(v)\bx\cdot\hat\bn_i\geq 0.\nonumber
\end{align}
Combining these conditions yields  $\sigma_v\sinh^2\rho_v\geq 0$,
which contradicts \eqref{sigcond}. Hence, blueshifts cannot occur even near the initial singularity of evolving spacetimes. 

{\bf Relation to the Wilson loop observables}

Formula \eqref{redshift} provides a simple expression for the redshift in terms of  the $ISO^+_0(2,1)$-valued holonomies along the elements of the fundamental group $\pi_1(M)\cong\Gamma$, the parameters specifying the observer's worldline and of the observer's eigentime at the emission of the lightray.  As in the previous examples, one finds that Wilson loop observables arise naturally in this description.  For observers whose momentum vector becomes parallel to the worldline of a point on the axis of an element
$v\in\Gamma$ in the limit $t\rightarrow\infty$,  the parameters $\rho_v$, $\nu_v$ are given by \eqref{msaxis}. For such an observer,  formula \eqref{redshift} thus establishes a direct relation between the redshift and the two fundamental Wilson loop observables associated to $v\in\Gamma$, the mass $m_v$ and the spin $s_v$ which characterise, respectively, the redshift in the limit $t\rightarrow\infty$ and the redshift near the initial singularity of the spacetime.  

We thus find that all of the three measurements considered in this section - the time intervals \eqref{nicesol} elapsed between the emission and reception of a returning lightray, the directions \eqref{direval} in which the light is sent to return and the angles \eqref{angledef} between these directions as well as the redshift \eqref{redshift} - are given explicitly as functions of the holonomy variables characterising the spacetime, of the observer's eigentime at the emission of the lightray and of the parameters which characterise his initial position and momentum. 

Moreover,  the expressions for these functions are simple and have a direct geometrical  interpretation. They are related to the lengths of closed spacelike geodesics in the  static spacetimes associated with $\Gamma$ and to the deflection of the corresponding  geodesics on the grafted strips in the evolving spacetimes. In all cases, the two canonical  Wilson loop observables associated with $v\in\Gamma$ characterise the measurements of a special set of observers - those whose momentum vector becomes parallel to the worldline of a point on the axis of 
$v\in\Gamma$ in the limit $t\rightarrow\infty$.


\section{Phase space, time and geometry}
\label{physdiscsect}

We are now ready to address the conceptual questions associated with the measurements considered in the previous section. In this section, we clarify the relation between these measurements and the physical phase space of the theory and demonstrate that specifying an observer with respect to the spacetime's geometry amounts to a gauge fixing procedure. We discuss the role of time in these measurements and show how the observer  can use them to reconstruct the full geometry of the spacetime.

\subsection{Gauge fixing, observables and the role of time}
\label{pobssect}

As shown in Sect.~\ref{messect},  all of the quantities under consideration,   the time intervals \eqref{nicesol} between the emission and reception of a returning lightray, the deflection angles \eqref{angledefl}, the angles \eqref{angledef} between the directions associated with returning lightrays  and the redshift \eqref{redshift}  are invariant under the global Poincar\'e transformations \eqref{globpoinc} which act simultaneously on the domain $D\subset \MM^3$, on the geodesics in $D$ which characterise the observer's worldline and on the holonomies.   In particular, this implies that the measurements are invariant under a shift \eqref{timeshift} of the observer's time origin which reflects the redundancy in the parametrisation of his worldline.
In that sense, the measurements are fully gauge invariant. 

However,  these measurements are  not  given as functions on the physical phase space \eqref{phspace} of the theory, which is parametrised by the holonomies modulo simultaneous conjugation. Without the specification of an observer, they are functions on the extended phase space $\text{Hom}_0(\pi_1(M), ISO^+_0(2,1))$, i.~e.~of the holonomies along a set of generators of the fundamental group, which depend on additional parameters, the observer's eigentime $t$ and the vectors $\bx\in H_1$, $\bx_0\in D$ which parametrise the observer's  worldline. They are invariant under a Poincar\'e transformation  which acts simultaneously on the holonomies and of the vectors $\bx\in H_1,\bx_0\in D$, but not under Poincar\'e transformations acting only on the holonomies.

However, as demonstrated in the previous section,  measurements associated with {\em specific} observers can be expressed in terms of the variables that parametrise the physical phase space of the theory. For observers whose momentum vector becomes parallel to the worldline of a point on the axis of $v\in\Gamma$ in the limit $t\rightarrow\infty$, the measurements \eqref{nicesol}, \eqref{angledefl}, \eqref{angledef} and \eqref{redshift} are given as functions of the eigentime and of the two fundamental Wilson loop observables \eqref{wabstr}, \eqref{massspindef} associated with $v\in\Gamma\cong\pi_1(M)$. This reflects a general pattern: in order to obtain functions on  the physical phase space, one needs to specify the worldline of the  observer by relating it to the holonomies.

\subsubsection{Specification of observers and gauge fixing}

 As the spacetimes do not contain any matter or distinguished reference frames at spatial infinity, the only physically meaningful way  of specifying an observer is with respect to the geometry of the spacetime itself. As this geometry is given uniquely by the holonomies $h(v)$, $v\in\Gamma$ in \eqref{grouphom}, this amounts to relating the vectors $\bx\in\hyp$, $\bx_0\in D$ parametrising the observer's worldline to the holonomies $h(v)$.

To specify the vector $\bx\in\hyp$ which determines the observer's reference frame  via \eqref{geodtime}, we need to select
 two elements of the Fuchsian group $v=\exp(n_v^aJ_a)$, $w=\exp(n_w^aJ_a)\in\Gamma\setminus\{1\}$ with respect to which we fix the observer's velocity vector $\bx$.  For instance, one can choose the generators $v_{a_1}$, $v_{b_1}$ in a fixed presentation  \eqref{fgroup} of $\Gamma$ which correspond to the $a$- and $b$-cycles of the first handle.  Alternatively, one could select   elements of $\Gamma$ whose traces are minimal, which amounts to selecting the two shortest geodesics on each static surface of constant cosmological time - if necessary specified uniquely by further conditions.

As explained in the appendix, 
the vectors $\bn_v,\bn_w$ that parametrise these group elements via the exponential map each define a unique geodesic in hyperbolic space $H_1\cong M_1^s$, the axes of $v$, $w\in\Gamma$. The axes of $v, w\in\Gamma$ intersect if and only if the wedge product  $\bn_v\wedge\bn_w$ is timelike. In that case their intersection point is given by the timelike unit vector
\begin{align}
\label{geodint}
\bx=\frac{\hat\bn_v\wedge\hat\bn_w}{\sqrt{1-(\hat\bn_v\hat\bn_w)^2}}\in H_1.
\end{align}
Otherwise, there is a unique point on the axis of $v$ whose geodesic distance from the axis of $w$ is minimal. It is given by the timelike unit vector
\begin{align}
\label{geodnonint}
\bx=\frac{\hat\bn_w - (\hat\bn_w\hat\bn_v)\hat\bn_v}{\sqrt{(\hat\bn_v\hat\bn_w)^2-1}}\in H_1.
\end{align}
 Identifying the vector $\bx\in\hyp$, which defines the reference frame of the observer, with these vectors thus amounts to fixing his direction of motion with respect to the geometry of the spacetime. It selects an observer whose momentum vector in the limit $t\rightarrow\infty$ becomes parallel to either the worldline of the intersection point of the axes of $v,w\in\Gamma$ or  of the point on the axis of $v\in\Gamma$ whose geodesic distance from the axis of $w\in\Gamma$ is minimal. 

To eliminate the remaining freedom in the choice of the observer, we need to specify the vector $\bx_0\in D$ which gives the observer's position at eigentime $t=0$. One possibility is to fix this initial position in such a way that the observer's eigentime coincides with the cosmological time. This amounts to selecting a point $\bp\in D$ in the regular domain for which the value of the Gauss map in \eqref{pointparam} coincides with the vector $\bx$ fixed via \eqref{geodint} or \eqref{geodnonint}, $N(\bp)=\bx$, and letting the initial position vector coincide with the retraction map $\bx_0=r(\bp)$. (In the case where the point $\bp$ lies on a strip glued in via the grafting construction one also needs to fix its distance from the edges of the strip.) This is a well-defined prescription, but its relation to the holonomy variables is complicated and implicit. 

We therefore choose a different prescription and specify the initial position of the observer with respect to the translation components of the holonomies  $h(v)$, $h(w)$.
From  \eqref{adrep} and \eqref{zparam2}  it follows that  for any $v=\exp(n_v^aJ_a)\in\Gamma$ it is  possible to fix $\bx_0\in D$ in such a way that
\begin{align}
\label{translfix}
h(v)g_{x,x_0}(0)-g_{x,x_0}(0)=\alpha_v\cdot \hat\bn_v\qquad \alpha_v\in\RR.
\end{align}
The residual freedom in the choice of the vector $\bx_0$ consists of  translations in the direction of $\bn_v$, which can be fixed by requiring that the parameter $\tau_{w}$ vanishes for some $w=\exp(n_w^aJ_a)\in\Gamma\setminus \{1\}$ chosen such that $\bn_v$, $\bn_w$ are linearly independent.

Hence, the observer can be specified uniquely via the following prescription: one selects two non-trivial elements in the fundamental group $\lambda,\xi\in\pi_1(M)$ and fixes the observer's velocity vector $\bx$  according to \eqref{geodint} if the axes of the associated elements $v_\lambda$, $v_\xi\in\Gamma$ intersect and according to \eqref{geodnonint} otherwise. One then fixes the observer's initial position $\bx_0$ by imposing the condition \eqref{translfix} for $\lambda$
and the additional condition $\tau_{\xi}=0$. Note that such a choice for $\bx_0$ can imply that $\bx_0$ now lies outside of the domain $D\in \MM^3$. However, in that case there still exists a time $t_0\in\RR$ such that $g_{x,x_0}(t)\in D$ $\forall t>t_0$. The situation therefore corresponds to an observer who has chosen the origin  of his eigentime prior to the eigentime at which it came into existence. It can be remedied via a  shift \eqref{timeshift} of the observer's time origin. 

With this specification of the observer, the parameters $\rho_v$, $\sigma_v$, $\tau_v$, $\nu_v$ in  formulas \eqref{nicesol}, \eqref{angledefl}, \eqref{redshift} for, respectively,  the elapsed time, the angles and the redshift are given as conjugation invariant functions of the holonomies of a set of generators of the fundamental group $\pi_1(M)$ and hence as functions of the physical observables.
This provides an explicit expression for the eigentime \eqref{nicesol} elapsed between the emission and reception of a returning lightray, for the angles \eqref{angledef} between the directions in which the light is sent in order to return and for the redshift \eqref{redshift} as functions of the physical phase space
and of the  observer's eigentime $t$ at the emission for all returning lightrays.

It is important to distinguish the quantities \eqref{nicesol}, \eqref{angledefl}, \eqref{redshift},  which depend on the observer's worldline and are functions on the extended phase space $\text{Hom}_0(\pi_1(M), ISO^+_0(2,1))$,
from their gauge fixed counterparts on the physical phase space \eqref{phspace}, for which the observer is specified with respect to the geometry of the spacetime.  While the former constitute partial observables in the terminology of Rovelli \cite{rovobs1}, the latter are Dirac observables on the physical phase space. The distinction between the two quantities has important consequences for the associated quantum theory.  

As shown in  \cite{thombianc}, the spectra of the  quantum operators associated to partial and Dirac observables can differ fundamentally, which implies that one needs to be careful when deciding which of these operators should be interpreted as a physical measurement.
However, the examples studied in \cite{thombianc} have been criticised as artificial \cite{rovspec}. The measurements considered in this paper would allow one to investigate  this issue   for physically meaningful quantities with a clear interpretation.   By considering the associated operators in the quantum theory, one would obtain a framework in which  the spectra of  partial and complete observables could be investigated and interpreted in a relevant model of quantum gravity.

\subsubsection{The role of time}

From the discussion in the preceding sections it follows that the observer's eigentime $t$ enters the theory as additional parameter which itself is neither a function 
on the physical phase space nor a parameter with respect to which the physical states evolve. Rather, it establishes a relation between time-dependent quantities that can be measured by an observer and points in the physical phase space that characterise the  geometry of the spacetime. 

This role of the eigentime and observers in (2+1)-dimensional vacuum spacetimes provides an example of Rovelli's concepts of partial and complete observables \cite{rovobs1} and evolving constants of motion \cite{rovtime1,rovtime2}.  All of the quantities under consideration, the eigentime \eqref{nicesol} elapsed between the emission and reception of a returning lightray, the  angles \eqref{angledefl}, \eqref{angledef}  and the redshift \eqref{redshift} 
are physical  observables insofar as they  can be measured 
by an observer. 

However, neither of them is a function on the physical phase space. 
The Dirac observables which are functions on the physical phase space  \eqref{phspace} are  the {\em relations} between these quantities and the observer's eigentime $t$ at which the lightray is emitted, i.~e.~expressions \eqref{nicesol}, \eqref{angledefl}, \eqref{angledef} and \eqref{redshift} for these quantities {\em as functions} of the  emission time. Alternatively, one can consider the time intervals \eqref{nicesol}, the angles \eqref{angledef} and the redshift \eqref{redshift} for a fixed value of the emission time. 

The eigentime $t$ itself is not a function on phase space. To obtain a Dirac observable corresponding to a time interval, one needs to specify this time with respect to two events in the spacetime.   
For instance, one can consider the eigentime elapsed between two measurements of the return time \eqref{nicesol}, the angles \eqref{angledefl}, \eqref{angledef} or the redshift \eqref{redshift} which yield fixed values $c_1,c_2$. This amounts to setting the left hand side of equations  \eqref{nicesol}, \eqref{angledefl}, \eqref{angledef} or  \eqref{redshift} equal to the constants $c_1$, $c_2$ and solving them for the eigentime $t$. After subtracting the resulting values for $t$, one then obtains a function on the physical phase space \eqref{phspace}, which involves the constants $c_1,c_2$ as parameters. 

Another possibility is to consider a fixed eigentime $t$ and to compare two measurements for  returning lightrays associated with different geodesics. 
An example is  the time $\Delta t_{\lambda_2}$ elapsed between the emission and reception of a returning lightray sent in the direction corresponding to a geodesic $\lambda_2$ under the condition that the time $\Delta t_{\lambda_1}$ elapsed between the emission and reception of a returning lightray sent along another geodesic $\lambda_1$ has a fixed value $\Delta t_{\lambda_1}=c$.
 This amounts to solving the condition  $\Delta t_{\lambda_1}(t)=c$ given by equation \eqref{nicesol} for the emission time $t$  and substituting this value for $t$  back into the corresponding equation for $\lambda_2$. One obtains  a function that depends on the physical observables and expresses the return time for one lightray as a function of the other.  
 
 Alternatively, one can express the value of one measurement associated with a given returning lightray as  a function 
 of another.  For instance, solving equation \eqref{angledefl} for $t$ and substituting this value into  \eqref{nicesol} yields an expression for the eigentime elapsed between the emission and reception of a returning lightray in terms of the deflection angle 
 \begin{align}
 \label{relfunct}
 \Delta t(\varphi_v)=\tfrac { \nu_v} 2\tanh \tfrac {\rho_v} 2 \sqrt{\cot^2\varphi_v+4\cosh^2\tfrac {\rho_v} 2}.
 \end{align}
 
 This expression becomes ill-defined for $\varphi_v\rightarrow 0$, which is the case for $t\rightarrow\infty$ in evolving spacetimes and for all values of $t$ in the static  spacetimes. In these cases, the angle $\varphi_v$ becomes independent of the emission time and takes the same value for all lightrays. It therefore  does no longer encode the information necessary for determining the  time elapsed between the emission and reception of returning lightrays.
  
In all cases, one finds that the physical measurements  are relational observables. In other words, the holonomy variables and Wilson loops which parametrise the physical phase space of the theory do not
directly determine the measurements of the observer. Rather, they determine the {\em relation} between different measurements and allow him to express one measurable quantity as a function of another.  

This has direct implications for the investigation of conceptual questions of quantum gravity in the quantised theory.  To determine if the spectra 
 of physical operators corresponding to measurements by observers are discrete or continuous, one needs to  take into account that physical observables and hence the associated operators on the Hilbert spaces of quantum theory only encode the relation between such measurements and not the measurements themselves.  It is therefore meaningless to ask, for instance,  if time is discrete or continuous near the Planck scale. Rather, one should ask questions such as 
``Is the eigentime elapsed between emission and reception of lightray measured by a specific observer characterised in terms of the geometry of the spacetime discrete or continuous {\em as a function} of other variables measured by this observer?"  Moreover,  the measurements under consideration need to be chosen carefully, as naive choices such as \eqref{relfunct} diverge already on the classical level.

\subsection{Reconstruction of the  holonomies from the measurements of an observer}
\label{geomreconstruct}

In the previous sections we derived explicit expressions for the measurements by an observer in terms of the fundamental variables which parametrise the phase space of the theory. This raises the question if and how such an observer can use these measurements to determine the physical state of the system.

 As the physical phase space \eqref{phspace}  is parametrised by the holonomies along the elements of the fundamental group $\pi_1(M)\cong\Gamma$ modulo simultaneous conjugation \eqref{globpoinc}, this amounts to reconstructing these holonomies from the measurements of the time intervals \eqref{nicesol}, the directions and angles \eqref{direval}, \eqref{angledefl}, \eqref{angledef}  and the redshift \eqref{redshift}.  Moreover, this reconstruction of the geometry needs to take into account that beyond such measurements, the observer has no means of determining his initial position and his reference frame. As discussed in the previous subsection, specifying the observer's reference frame is physically meaningful only with respect to the geometry of the spacetime to be determined through these measurements. Hence, we have to assume in the following that the observer is ignorant of the vectors $\bx\in H_1$, $\bx_0\in D$ which characterise his worldline via \eqref{geodset}.

We start by considering the Lorentzian components of the holonomies \eqref{grouphom}. 
To find an explicit prescription that allows the observer to reconstruct these quantities
from his measurements, we consider formulas \eqref{nicesol} for the time intervals elapsed between the emission and reception of a returning lightray and formula \eqref{direval} for the directions in the limit  $t\rightarrow\infty$
\begin{align}
\label{tauform}
\lim_{t\rightarrow\infty} \frac d {dt} \Delta t(t,v, \bx,\bx_0)=e^{\rho_v}-1\qquad \lim_{t\rightarrow\infty}
\hat\bp_v(t,v,\bx,\bx_0)=\frac{\Pi_{\bx^\bot}(\Ad(v)\bx)}{|\Pi_{\bx^\bot}(\Ad(v)\bx)|},
\end{align}
where $\Pi_{\bx^\bot}$ is the projection on the orthogonal complement of the observers momentum vector. The parameter  $\rho_v$ is the geodesic distance between $\bx$ and $\Ad(v)\bx$ in $H_1$ and corresponds to the length of the closed geodesic associated with $v\in\Gamma$ on the static spatial surface $M_1^s\cong \hyp/\Gamma$.

After an arbitrary choice of a vector $\bx\in H_1$, these formulas allow the observer to determine both the directions of the geodesic from $\bx$ to all images $\Ad(v)\bx$ in $H_1$ and the distance of these images from $\bx$. Hence, after selecting an arbitrary vector $\bx\in H_1$, the observer can reconstruct all of its images $\Ad(v)\bx\in H_1$ under the action of the cocompact Fuchsian group $\Gamma$.  This allows the observer to reconstruct the Dirichlet region of $\Gamma$ and to obtain an explicit set of generators for $\Gamma$ as follows. 

As explained in the appendix, the Dirichlet region $R^\Gamma_D(\bx)$  is obtained as the set of points in $H_1$ whose geodesic distance from $\bx$ is less than or equal  to their geodesic distance from all images $\Ad(v)\bx$, $v\in\Gamma$. The observer  can thus reconstruct the Dirichlet region $R^\Gamma_D(\bx)$ by  considering all perpendicular bisectors of the geodesic segments joining $\bx$ and $\Ad(v)\bx$ in $H_1$ and intersecting the associated half-hyperboloids of points that lie on the same side of these bisectors as $\bx$.   The  result  is a $2k$-gon  $R^\Gamma_D(\bx)$, $k\geq 2g$, in which all of the $2k$ sides are geodesic arcs and the sides are identified pairwise by certain elements 
 $v_1,\ldots,v_{k}\in\Gamma$.  These group elements form a set of generators\footnote{Note that this set of generators is not necessarily of the form \eqref{fgroup}.} of $\Gamma$.
 
 By considering the time intervals \eqref{nicesol} between the emission and reception of a returning lightray and the directions \eqref{direval} in the limit $t\rightarrow\infty$ and for all returning lightrays, 
  the observer can thus reconstruct the cocompact Fuchsian group $\Gamma$ as well as a presentation in terms of a set of generators and relations. This amounts to determining the Lorentzian component of the holonomies which characterise  the geometry of the associated static spacetime approached in the limit $T\rightarrow\infty$. These measurements thus allow the observer to fully reconstruct the geometry of the spacetime for $T\rightarrow\infty$.

 Note that the choice of the basepoint in this construction does not affect the result.  A different choice of basepoint yields a set of generators that is related to the original set by global conjugation with an element $v_0\in SO^+_0(2,1)\cong PSU(1,1)$. This corresponds to a global Lorentz transformation acting on the domain $D\subset \MM^3$ and the holonomies according to \eqref{globpoinc} and is a gauge symmetry.  In other words, the observer does not need to know its direction of motion relative to the geometry of the spacetime in order to reconstruct the static spacetime approached in the limit $T\rightarrow\infty$. 

To determine the translational component of the holonomies \eqref{grouphom} up to global conjugation, the observer needs to measure the parameters $\sigma_v,\tau_v,\nu_v$ in \eqref{zparam2} for all $v\in\Gamma$ and then insert the value of his chosen basepoint $\bx$ and the group elements $v\in\Gamma$ into \eqref{zparam2}. This can be done, for instance, by considering how the directions into which the light needs to be emitted to return to the observer change with time.  Formula  \eqref{angledefl} for the deflection angle allows the observer to determine the parameter $\nu_v$
\begin{align}
\lim_{t\rightarrow\infty} \frac d {dt}\cot \varphi_v(t,v,\bx,\bx_0)=\frac{e^{\rho_v}+1}{\nu_v}\qquad \forall v\in\Gamma.
\end{align}
Knowing the values of the parameters $\rho_v$ and  $\nu_v$ for all elements $v\in\Gamma$, the observer can then reconstruct the quantity $(t+\sigma_v)$ by considering for instance, the frequency shift \eqref{redshift}. His measurements of the time intervals \eqref{nicesol} between the emission and reception of the returning lightrays allow him then to reconstruct the parameters $\tau_v$ for all $v\in\Gamma$.

Note that it is not possible for the observer to reconstruct the parameters $\sigma_v$ from his measurements unless he knows the origin of his eigentime.   This reflects  the invariance   under the shift \eqref{timeshift} of the observer's time origin which is due to the redundancy in the parametrisation \eqref{geodset} of his worldline.  It corresponds to a global Poincar\'e transformation \eqref{globpoinc} with $(v_0,\ba_0)=(1, t_0\bx)$, under which the parameters $\sigma_v$ transform as  $\sigma_v\rightarrow \sigma_v-t_0$ for all $v\in\Gamma$.  Hence, the observer needs to  specify an origin for his eigentime by selecting an arbitrary value for the parameter $\sigma_v$ for one of the group elements $v\in\Gamma\setminus\{1\}$.
Observers who make different choices for these parameters obtain holonomies which differ by a global Poincar\'e transformation \eqref{globpoinc} which reflects the gauge freedom of the theory. 

Together, the two steps of the construction allow the observer (assumed to be ignorant of the parameters characterising his own worldline) to determine the holonomies \eqref{grouphom} up to global conjugation. As these holonomies parametrise the physical phase space \eqref{phspace} of the theory and determine the geometry of the spacetime uniquely, we have thus demonstrated that a general observer in an evolving spacetime can reconstruct the full geometry of the spacetime from physical measurements analogous to gravitational lensing. Moreover, the results in this subsection provide an explicit algorithm for doing so.

\section{Outlook and conclusions}

\label{outlook}

In this paper we addressed the problem of relating  the gauge and diffeomorphism invariant observables that paramet\-rise the phase space of (2+1)-dimensional gravity to realistic physical measurements.  By considering an observer who probes the geometry of the spacetime by emitting returning lightrays, we identified several quantities that are  directly related to the observables of the theory and have a clear physical interpretation:  the eigentime elapsed between the emission and reception of a returning lightray, the directions into which the light needs to be sent in order to return to the observer as well as  the angles between them and the frequency shift between the emitted and the returning lightray. 

We derived explicit expressions for these measurements in terms of the variables that paramet\-rise the physical phase space of the theory and are the fundamental building blocks in its quantisation, the holonomies and Wilson loop observables.  More specifically, we found that the measurements performed by observers  are given as functions of the holonomies, the observer's eigentime at the emission of the lightray and of additional parameters that characterise the observer's worldline. We demonstrated that specifying an observer with respect to the geometry of the spacetime amounts to a gauge fixing prescription and that the associated measurements are Dirac observables, functions on the physical phase space that depend on the emission time as an additional parameter.

We discussed the physical interpretation of these measurements and analysed how they encode the geometry of the spacetime. This gave rise to an explicit prescription that allows an observer to reconstruct the values of the physical observables and hence the  physical state of the system from the results of his measurements. In particular, we showed that the fundamental gauge and diffeomorphism invariant observables of the theory, the Wilson loops associated with closed curves in the spacetime, arise naturally as parameters in  the measurements of a special set of observers. They are associated with observers whose momentum three-vector is parallel to the worldline of points on the associated geodesic in the limit where the cosmological time and the eigentime tend to infinity.

Our results thus provide a set of observables with a clear physical interpretation that are directly related to realistic physical measurements performed by observers in a spacetime. They also shed light on several conceptual questions of quantum gravity that manifest themselves in the description. In particular, they  serve as a concrete example which allows one to investigate the role of time in the theory and the relation between partial  and complete  observables. 

This offers the prospect of defining and investigating the associated operators in the quantised theory.
The results of Sect.~\ref{messect} and Sect.~\ref{physdiscsect} could be  adapted in a straightforward manner to a formulation of the theory based on graphs or spin network functions.
They also could be generalised to vacuum spacetimes in Lorentzian (2+1)-gravity with a non-trivial cosmological constant, although the calculations will be more involved and additional complications can be expected for the de Sitter case\footnote{In that case the holonomies (modulo global conjugation) do not determine the geometry of the spacetime uniquely \cite{mess,npm}. Rather, there is an infinite discrete set of spacetimes for each value of the holonomy variables.}.

The application of the results to the quantised theory would allow one to investigate fundamental questions of quantum gravity in a concrete and well-defined example. In particular, it could be used to investigate the role of time in the quantum theory and to address questions about  the spectra of physical operators, which have been subject to much debate in the quantum gravity community  \cite{thombianc, rovspec}. 

Finally,  the results of this paper might also be relevant to cosmological applications of quantum
gravity in (3+1) dimensions. The measurements investigated in this paper are similar to those in gravitational lensing and the description could easily be generalised to include external sources. 
Moreover, the spacetimes considered in this paper have realistic physical properties such as initial singularities and expansion with the cosmological time and share many features with the Bianchi spacetimes studied extensively in (loop) quantum cosmology. It therefore seems plausible that the results of this paper would have counterparts and analogies in that context.


\section*{Acknowledgements}
I am grateful to Franceso Bonsante for extensive discussions during my  visit to Pavia and to the Department of Mathematics, University of Pavia, for hospitality and for supporting this visit. I thank Louis Crane for discussions about gravitational lensing, Bianca Dittrich for discussions about observables and for comments on a draft of this paper and Frank Hellmann, Jorma Louko and Andrei Starinets for comments on the draft of this paper.
 
The research was partly undertaken at the Perimeter Institute for Theoretical Physics and partly at the University of Nottingham. Research at Perimeter Institute is supported by the Government
of Canada through Industry Canada and by the Province of Ontario through
the Ministry of Research \&
 Innovation. The work at the University of Nottingham was supported by the Marie Curie Intra-European Fellowship PIEF-GA-2008-220480.


\appendix

\section{Hyperbolic geometry and Fuchsian groups}
\label{hypgeomfuchs}

In this appendix we summarise some notions from two-dimensional hyperbolic geometry and the theory of Fuchsian groups required for the understanding of this paper. For a more thorough treatment of hyperbolic geometry and Riemann surfaces we refer the reader to the books by Benedetti and Petronio \cite{bp} and by Farkas and Kra \cite{farkkra}.  An accessible introduction to the theory of Fuchsian groups is given in the book \cite{fgroups} by Katok. 

\subsection{Two-dimensional hyperbolic geometry}
\label{hypgeom}

\subsubsection*{Poincar\'e disc and hyperboloid model}

Two-dimensional hyperbolic geometry is concerned with the geometry of two-dimensional hyperbolic space $\HH^2$, which can be realised either as the Poincar\'e disc, the upper half-plane or a hyperboloid in Minkowski space. In this appendix, we focus on the disc and the hyperboloid model. The Poincar\'e disc model is given by\begin{align}
\label{unitdisc}
D=\{z\in\CC\;|\; |z|^2<1\}\qquad g_D=\frac{4dzd\bar z}{(1-|z|^2)^2}.
\end{align}
The hyperboloid model is the unit hyperboloid in Minkowski space $\MM^3$ with the  metric induced by the Minkowski metric
\begin{align}
\label{hypmodel}
H_1=\{\bx\in\MM^3\;|\; \bx^2=-1,\, x^0>0\}\qquad g_{H_1}=\eta|_{H_1}.
\end{align}
The two models are isometric, with the identification of a 
 point $\bx=(x^0,x^1,x^2)\in H_1$ and a point $z=z_1+i z_2$ on the Poincar\'e disc  given by
\begin{align}
\label{hypdisc}
z=\frac{x^1+ix^2}{1+x^0}\qquad x^0=\frac{1+|z|^2}{1-|z^2|}\qquad x^1+ix^2=\frac{2z}{1-|z|^2}.
\end{align}

\subsubsection*{Geodesics}

Geodesics of the Poincar\'e disc are straight lines through the origin and circles which intersect its  boundary  $\partial D=\{z\in\CC: |z|=1\}$ orthogonally. They are characterised by the equations
\begin{align}
\label{geodequ}
2\text{Re}(z\bar w)=0 \qquad\qquad |z|^2+1=2\text{Re} (z\bar w),
\end{align}
where $w$ is, respectively,  a vector orthogonal to the line  or  the centre of the circle. For any two points $p,q\in D$ there is a unique geodesic $c_{p,q}:[0,1]\rightarrow D$ with $|\dot c|=1$, $c(0)=p$ and $c(1)=q$. This geodesic  has a unique perpendicular bisector $c^\bot_{p,q}$ given by \eqref{geodequ} with
\begin{align}
\label{perpcenter}
w=\frac{(1-|p|^2)\bar q-(1-|q|^2)\bar p}{|q|^2-|p|^2}.
\end{align}
The geodesic distance of two points $w,z\in D$ is defined as the infimum of the lengths of  piecewise smooth curves connecting $w$ and $z$
\begin{align}
\label{hypdistdef}
\rho(z,w)=\inf\{l(c)\;|\; c:[0,1]\rightarrow D, c(0)=z, c(1)=w\}\quad l(c)=\int_0^1 \sqrt{g_D(\dot c, \dot c)} dt.
\end{align}
It is the length of the (unique) geodesic  connecting $w$ and $z$ and is given by
\begin{align}
\label{hypdist}
\sinh \frac{\rho(z,w)} 2=\frac{|z-w|}{(1-|z|^2)(1-|w|^2)}.
\end{align}
The perpendicular bisector  \eqref{perpcenter} is the set of points   equidistant from the two points $p,q\in D$ 
\begin{align}
\label{perpdist}
c^\bot_{p,q}=\{z\in D\;|\; \rho(z,p)=\rho(z,q)\}.
\end{align}
In the hyperboloid model, geodesics are given as the intersections of the hyperboloid  $H_1$ and planes $\bn^\bot$ through the origin with spacelike normal vectors $\bn$  as shown in Fig.~\ref{grafting} a)
\begin{align}
\label{plane}
\bn^\bot=\{\bx\in\RR^3\;|\; \bx\cdot\bn=0\}\qquad \bn^2>0.
\end{align}
The unique geodesic through two points $\bx,\by\in H_1$ is the intersection of $H_1$ with the plane with normal vector $\bx\wedge\by$. Using formula \eqref{hypdisc} for the identification of the disc with the hyperboloid, one finds that the geodesic distance \eqref{hypdistdef} of two points $\bx,\by\in H_1$  is given by
\begin{align}
\label{hypdistmink}
\cosh\rho({x,y})=-\bx\cdot \by.
\end{align}

\subsubsection*{Isometries}
The isometry group of two-dimensional hyperbolic space $\HH^2$ is the group
\begin{align}
PSU(1,1)=SU(1,1)/\mathbb Z_2\cong PSL(2,\RR)=SL(2,\RR)/\mathbb Z_2\cong SO(2,1)^+_0.
\end{align}
Its action on the Poincar\'e disc $D$ is given by its $SU(1,1)$ representation 
\begin{align}
\label{isomact}
v=\left(\begin{array}{cc} a & c\\ \bar c & \bar a\end{array}\right)\quad |a|^2-|c|^2=1\;:\qquad  z\mapsto\frac{az+\bar c}{cz+\bar a}.
\end{align}
As this action on the Poincar\'e disc  is invariant under $v\rightarrow -v$, it induces an action of $PSU(1,1)=SU(1,1)/\mathbb Z_2$ on $D$. Elements of $SU(1,1)$ are called hyperbolic, parabolic and elliptic, respectively, if $|\tr(v)|>2$, $|\tr (v)|=2$, $|\tr(v)|<2$. Hyperbolic elements have two fixed points on $\partial D$, parabolic elements a single fixed point in $\partial D$ and elliptic elements a single fixed point in $D$. 

The axis of a hyperbolic element $v\in PSU(1,1)$ is the unique geodesic through its two fixed points and is mapped to itself by the action of $v$ in \eqref{isomact}. It is given by the equation
\begin{align}
\label{axis}
\text{Im}(a)(1+|z|^2)=2\text{Im}(cz).
\end{align}
In the hyperboloid model, the isometry group $PSU(1,1)\cong SO(2,1)^+_0$ acts via its $SO(2,1)$-representation which agrees with its adjoint representation on $\mathfrak{su}(1,1)\cong\mathfrak{so}(2,1)\cong \RR^3$  given by \eqref{adjact}. Using the parametrisation \eqref{adrep} in terms of a vector $\bn\in\RR^3$ via the exponential map, one finds that elements are hyperbolic, parabolic and elliptic, respectively, if  $\bn^2>0$ (spacelike), $\bn^2=0$ (lightlike) and $\bn^2<0$ (timelike). The axis of a hyperbolic element parametrised as in \eqref{adrep} is the intersection of the hyperboloid $H_T$ with the plane $\bn^\bot$.


\subsection{Cocompact Fuchsian groups}
\label{fuchsgroup}

A Fuchsian group is a discrete subgroup of  $PSU(1,1)\cong PSL(2,\RR)\cong SO(2,1)^+_0$.  A cocompact Fuchsian group of genus $g$ is a Fuchsian group $\Gamma$ such that the quotient $\HH^2/\Gamma$ is a compact orientable surface of genus $g$. It has a presentation in terms of 
 $2g$ generators and a defining relation \begin{align}
\label{fgroupdef}
\Gamma=\langle a_1,b_1,\ldots, a_g,b_g \in PSU(1,1)\;|\; [b_g, a_g^\inv]\cdots[b_1, a_1^\inv]=1\rangle,
\end{align}
where $[u,v]=u v u^\inv v^\inv$ is the group commutator.
All (non-unit) elements of a  cocompact Fuchsian group are hyperbolic. Its action  on $\HH^2$ via \eqref{isomact} is free and properly discontinuous, which implies that the quotient 
 $\Sigma_g=\HH^2/\Gamma$
is a two-dimensional manifold of genus $g$  with a metric of constant curvature -1 induced by the metric on $\HH^2$.  The geodesics on the surface $\Sigma_g$ are the images of $\Gamma$-equivalence classes of geodesics on $\hyp$ under the projection $\hyp\rightarrow\hyp/\Gamma$.  

A fundamental region for a Fuchsian group $\Gamma$ is a closed region $F\subset\hyp$ such that
\begin{align}
\label{tess}
\bigcup_{v\in\Gamma} vF=\hyp\qquad\quad\text{and}\quad \qquad \check{(vF)}\cap\check{F}=\emptyset\;\forall v\in\Gamma\setminus\{1\},\;\text{where}\; \check F=F\setminus\partial F.
\end{align}
Each fundamental region for $\Gamma$ induces a tessellation  of $\hyp$ via the action \eqref{isomact} of $\Gamma$. An example of a fundamental region is the Dirichlet region $R^\Gamma_D(z)$ based at a point $z\in D$, which is the set of points whose geodesic distance \eqref{hypdist} from $z$ is less than or equal to  their geodesic distance from all images of $z$ under the action of  $\Gamma$
\begin{align}
\label{dirreg}
R^\Gamma_D(z)=\{w\in D\;|\; \rho(z,w)\leq\rho(vz,w)\quad\forall v\in\Gamma\}.
\end{align} 
One can show that the Dirichlet region $R^\Gamma_D(z)$ is given as the intersection of the "half-planes"
 \begin{align}
 \label{dirhalfplane}
R^\Gamma_D(z)=\bigcap_{v\in\Gamma\setminus \{1\}} H_z(v)\qquad  H_z(v)=\{w\in D\;|\;\ \rho(z,w)\leq \rho(vz, w)\}.
 \end{align}
Due to the invariance of the hyperbolic distance $\rho$ under the isometry group $PSU(1,1)$, the associated tessellation \eqref{tess} takes the form
\begin{align}
\label{dirtess}
D=\bigcup_{v\in\Gamma} v\, R^\Gamma_D(z)=\bigcup_{v\in\Gamma} R^\Gamma_D(vz).
\end{align}
For a cocompact Fuchsian group of genus $g$, the Dirichlet region $R^\Gamma_D(z)$ is a compact, convex, connected region in $D$. Its boundary $\partial R^\Gamma_D(z)$ is the union of $2k\geq 4g$  geodesic arcs. These arcs
 are given as the perpendicular bisectors \eqref{perpdist} of certain geodesic segments $[z, vz]$, $v\in\{v_1^{\pm 1},\ldots, v_k^{\pm 1}\}\subset \Gamma$  for a finite number of elements  of $\Gamma$ and their inverses. Hence, the elements $v_1,\ldots,v_k\in\Gamma$ identify the sides of the Dirichlet region pairwise and form  a set of generators\footnote{Note that this does not in general  imply  $k=2g$ and neither that this set of generators is of the form \eqref{fgroupdef}. In the generic situation one has $k>2g$ and the generators satisfy a different set of relations. }  of $\Gamma$.
\begin{figure}[h]
\centering
\includegraphics[scale=0.4]{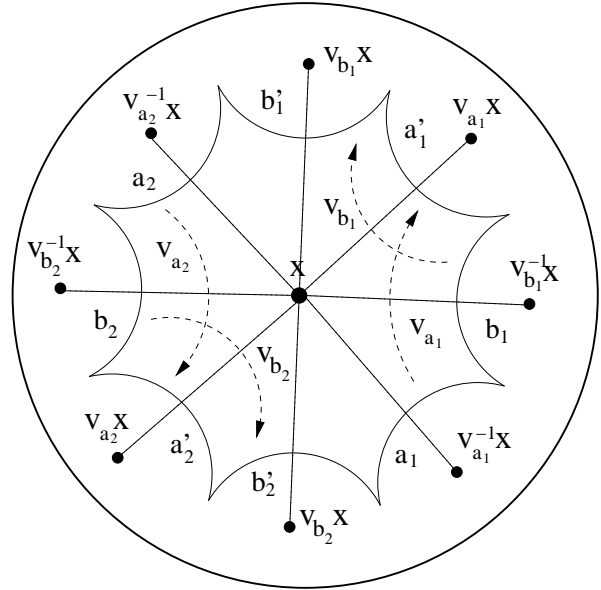}
\caption{\small{The standard fundamental region for a cocompact Fuchsian group of genus $g=2$.}}
\label{fundpoly}
\end{figure}
\begin{figure}[h]
\centering
\includegraphics[scale=0.3]{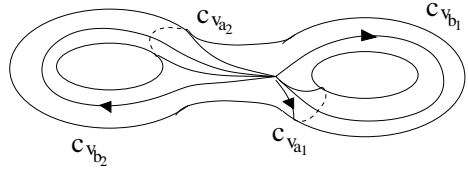}
\caption{\small{A set of curves representing the standard generators of the fundamental group of a genus 2 surface. The curves correspond to the dotted lines connecting the point $\bx$ to its images in Fig.~\ref{fundpoly}.}}
\label{fundgr}
\end{figure}

 It has been shown by Poincar\'e \cite{poinc} that starting from the Dirichlet region, it is possible to construct another fundamental region for $\Gamma$, the so called standard or canonical fundamental region $R^\Gamma_s$ of $\Gamma$. This is again a compact, convex, connected region in $D$ bounded by geodesic arcs which are identified pairwise by certain elements  of $\Gamma$. However, in this case the number of arcs is always $4g$. The geodesic arcs in its boundary are identified as shown in Fig.~\ref{fundpoly}, and the associated group elements $v_{a_1}$, $v_{b_1}$, ..., $v_{a_g}$, $v_{b_g}\in\Gamma$  form a set of generators of $\Gamma$ as in \eqref{fgroupdef}. The quotient surface $\Sigma_g=\hyp/\Gamma$ is obtained by gluing the sides of the standard polygon pairwise as shown in Fig~\ref{fundpoly}. A set of curves representing the associated  generators of its fundamental group is depicted in Fig.~\ref{fundgr}.

\end{document}